\documentclass[pra,aps,showpacs,groupedaddress,superscriptaddress,twocolumn,toc=flat,biblatex,footinbib]{revtex4-1}
\usepackage{natbib}
\usepackage[table]{xcolor}
\usepackage[colorlinks=false]{hyperref}
\usepackage[utf8]{inputenc}
\usepackage{color}
\usepackage{bbm} 
\usepackage[english]{babel}
\usepackage{graphicx}
\usepackage{placeins}
\usepackage{multirow}
\usepackage{physics}
\usepackage{csquotes}
\usepackage{amsfonts,amsmath,amssymb,stmaryrd}
\usepackage{graphicx}
\usepackage{subfigure}  
\usepackage{epsfig}
\usepackage{mathrsfs}
\usepackage{verbatim}
\usepackage{centernot}
\usepackage{ulem}
\usepackage{longtable}
\usepackage{wasysym}
\usepackage{nicefrac}
\usepackage[framemethod=tikz]{mdframed}
\usepackage{array}
\usepackage{bm}	
\usepackage{cancel,ifthen}

\newcommand{\cmnt}[2][NoInPuT]{\ifthenelse{\equal{#1}{NoInPuT}}{}{{\color{red}\sout{#1}}} {\color{blue} #2}}
\renewcommand{\vec}[1]{\bm{#1}}

\LTcapwidth=\textwidth

\begin{document}

\normalem	

\title{Neural network approach to quasiparticle dispersions in doped antiferromagnets}

\author{Hannah Lange}
\affiliation{Ludwig-Maximilians-University Munich, Theresienstr. 37, Munich D-80333, Germany}
\affiliation{Max-Planck-Institute for Quantum Optics, Hans-Kopfermann-Str.1, Garching D-85748, Germany}
\affiliation{Munich Center for Quantum Science and Technology, Schellingstr. 4, Munich D-80799, Germany}

\author{Fabian Döschl}
\affiliation{Ludwig-Maximilians-University Munich, Theresienstr. 37, Munich D-80333, Germany}
\affiliation{Munich Center for Quantum Science and Technology, Schellingstr. 4, Munich D-80799, Germany}

\author{Juan Carrasquilla}
\affiliation{Department of Physics, 60 Saint George St., University of Toronto, Toronto, Ontario, M5S 1A7, Canada}
\affiliation{Vector Institute, MaRS Centre, Toronto, Ontario, M5G 1M1, Canada}
\affiliation{Department of Physics and Astronomy, University of Waterloo, Ontario, N2L 3G1, Canada}

\author{Annabelle Bohrdt}
\affiliation{Munich Center for Quantum Science and Technology, Schellingstr. 4, Munich D-80799, Germany}
\affiliation{University of Regensburg, Universitätsstr. 31, Regensburg D-93053, Germany}

\pacs{}

\date{\today}

\begin{abstract}
Numerically simulating spinful, fermionic systems is of great interest from the perspective of condensed matter physics. However, the exponential growth of the Hilbert space dimension with system size renders an exact parameterization of large quantum systems prohibitively demanding. This is a perfect playground for neural networks, owing to their immense representative power that often allows to use only a fraction of the parameters that are needed in exact methods. Here, we investigate the ability of neural quantum states (NQS) to represent the bosonic and fermionic $t-J$ model -- the high interaction limit of the Fermi-Hubbard model -- on different 1D and 2D lattices. Using autoregressive recurrent neural networks (RNNs) with 2D tensorized gated recurrent units, we study the ground state representations upon doping the half-filled system with holes. Moreover, we present a method to calculate dispersion relations from the neural network state representation, applicable to any neural network architecture and any lattice geometry, that allows to infer the low-energy physics from the NQS. To demonstrate our approach, we calculate the dispersion of a single hole in the $t-J$ model on different 1D and 2D square and triangular lattices. Furthermore, we analyze the strengths and weaknesses of the RNN approach for fermionic systems, pointing the way for an accurate and efficient parameterization of fermionic quantum systems using neural quantum states.

\end{abstract}

\maketitle


The simulation of quantum systems has remained a persistent challenge until today, primarily due to the exponential growth of the Hilbert space, making it exceedingly difficult to parameterize the wave functions of large systems using exact methods. Since the seminal work of Carleo and Troyer \cite{Carleo2017}, the idea of using neural networks to simulate quantum systems \cite{Carleo2017,Torlai2016,Torlai2018,Torlai2019,Carrasquilla2019} has been applied successfully for a large number of quantum systems, leveraging various neural network architectures. These architectures include restricted Boltzmann machines \cite{Torlai2016,Torlai2018}, convolutional neural networks (CNNs) \cite{Schmale2022}, group CNNs \cite{roth2021}, autoencoders \cite{Rocchetto2018} as well as autoregressive neural networks such as recurrent neural networks (RNNs) \cite{Morawetz2021,HibatAllah2020,Sharir2020,Luo2021,Luo2022}, with neural network representations of both amplitude and phase distributions of the quantum state under consideration. These neural quantum states (NQS) make use of the innate ability of neural networks to efficiently represent probability distributions. When applying them to represent quantum systems, this ability can help to reduce the number of parameters required to encode the system.

Despite their representative power, NQS have been shown to face challenges during the training process, for example when they are trained to minimize the energy, i.e. to represent ground states. This results from the intricate nature of the loss landscape, characterized by numerous saddle points and local minima that complicate the search for the global minimum \cite{Bukov2021}. One promising avenue to overcome this problem is the use of many uncorrelated samples during the training. This strategy is facilitated when using autoregressive neural networks \cite{Uria2016, Humeniuk2022}, allowing to directly sample from the wave functions' amplitudes. Autoregressive networks have already been applied in the physics context \cite{Carrasquila2021, Wu2019}, such as for variational simulation of spin systems \cite{HibatAllah2020,Sharir2020, Luo2022,Luo2021}.  

Many works have so far focused 
on NQS representations of spin systems at half-filling, revealing that NQS can be used to study a variety of phenomena that are relevant to state-of-the-art research, as e.g. shown for RNN representations on various lattice geometries, including frustrated  spin systems \cite{HibatAllah2022,HibatAllah2020}, and systems with topological order \cite{Hibatallah2023}. For all of these systems, the physics becomes even richer when introducing mobile impurities, e.g. holes, into the system, yielding a competition between the magnetic background and the kinetic energy of the impurity. Simulating such systems holds particular relevance for understanding high-temperature superconductivity, where the superconducting dome arises upon doping the antiferromagnetic half-filled state with holes \cite{Keimer:2015}. The search for NQS that are capable of representing such spinful fermionic systems is still in its early stages. In recent years, first NQS have been developed that obey the fermionic statistics, simulating molecules \cite{Pfau2020,Spencer2020,Barret2022}, spinless fermions \cite{Humeniuk2022} and spinful fermions \cite{Nomura2017,Inui2021,Luo2019,Moreno2022}. Among those architectures are FermiNet \cite{Pfau2020,Spencer2020}, Slater-Jastrow ansätze \cite{Humeniuk2022,Nomura2017,Luo2019,Moreno2022} or variants of Jordan-Wigner transformations \cite{Choo2020,Barret2022,Yoshioka2021,Hermann2020,Inui2021}.\\

Here, we use an autoregressive neural network architecture, supplemented with a Jordan-Wigner transformation, to simulate ground states of the high interaction limit of the Fermi-Hubbard model, believed to capture essential features of high-temperature cuprate superconductors. Specifically, we use RNNs, proven to successfully model spin systems \cite{Morawetz2021,HibatAllah2020,HibatAllah2022,Hibatallah2023,Czischek2022,Moss2023}, and simulate the ground states of the fermionic (bosonic) $t-J$ model, both in one and two dimensions. In its more generalized form, known as the fermionic (bosonic) $t-$XXZ model, with anisotropic superexchange interactions denoted as $J_z$ and $J_\pm$, the Hamiltonian under consideration reads as follows:
\begin{align}
        \mathcal{H}_{tXXZ}=&-t \sum_{\langle \vec{i},\vec{j}\rangle, \sigma}\mathcal{P}_G\left( \hat{c}^{\dagger}_{\vec{i},\sigma} \hat{c}_{\vec{j},\sigma}+\mathrm{h.c.}\right) \mathcal{P}_G\notag
        \\&+J_z\sum_{\langle\vec{i},\vec{j}\rangle}\left(\hat{S}^z_{\vec{i}}\cdot \hat{S}^z_{\vec{j}}-\frac{1}{4}  \hat{n}_{\vec{i}}\hat{n}_{\vec{j}}\right) \notag \\
        & +J_{\pm}\sum_{\langle\vec{i},\vec{j}\rangle}\frac{1}{2}\left(\hat{S}^+_{\vec{i}}\cdot \hat{S}^-_{\vec{j}}+\hat{S}^-_{\vec{i}}\cdot \hat{S}^+_{\vec{j}}\right),
        \label{eq:tJ}
\end{align}
with the fermionic (bosonic) creation and annihilation operators $\hat{c}^{\dagger}_{\vec{i},\sigma}$ and $\hat{c}_{\vec{i},\sigma}$ for particles at site $\vec{i}$ with spin $\sigma$; spin operators are denoted by $\hat{\vec{S}}_{\vec{i}}=\sum_{\sigma,\sigma^\prime}\hat{c}^{\dagger}_{\vec{i},\sigma} \frac{1}{2}\vec{\sigma}_{\sigma \sigma^\prime}\hat{c}_{\vec{i},\sigma^\prime} $ as well as density operators by $\hat{n}_{\vec{i}}$ \cite{Auerbach2012}. For $J_z=J_{\pm}$, Eq. \eqref{eq:tJ} reduces to the  $t-J$ model and for $J_{\pm}=0$ to the $t-J_z$ model. \\

\begin{figure}[t]
\centering
\includegraphics[width=0.5\textwidth]{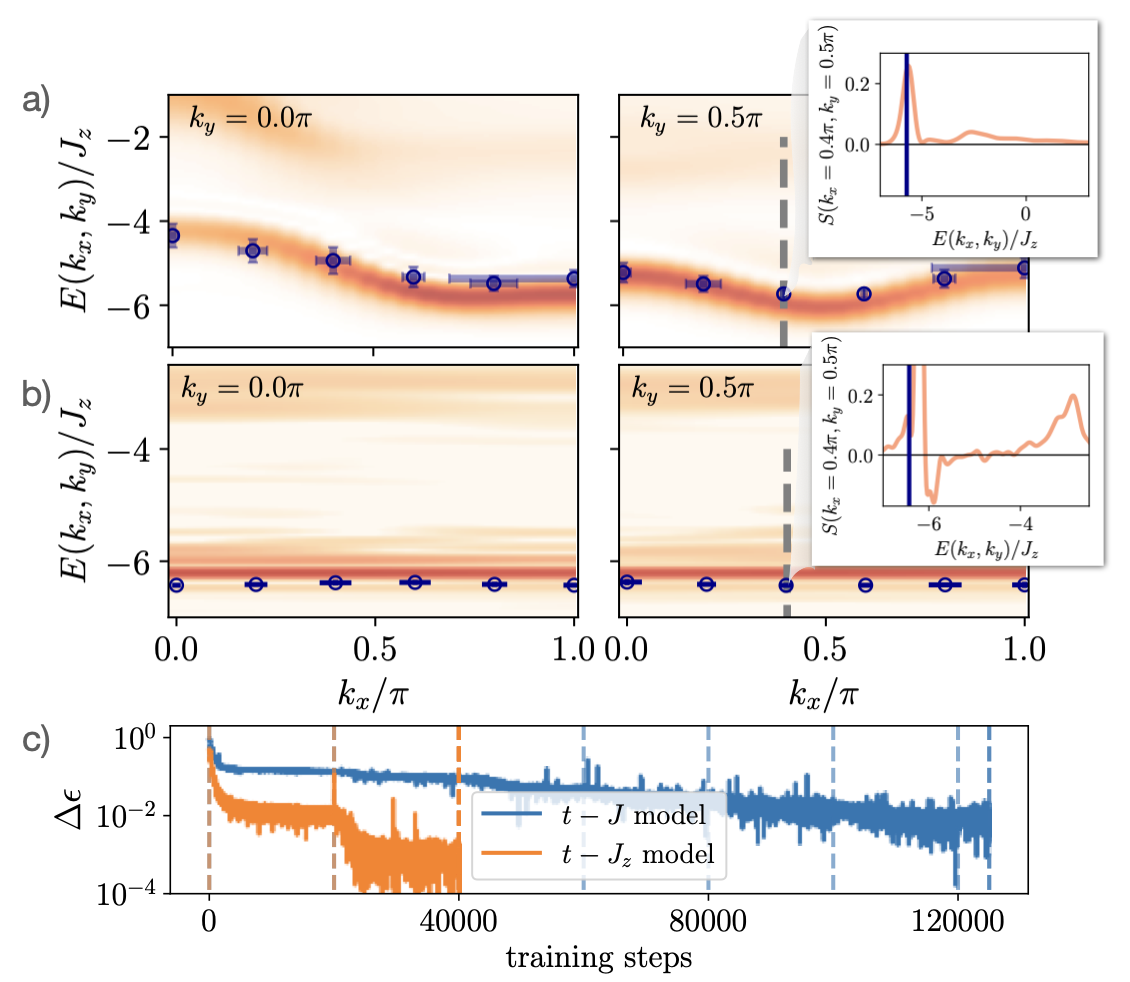}
\caption{$t-J$ and $t-J_z$ square lattice with $10\times 4$ sites, $t/J_z=3$ and open boundaries in $x$, periodic boundaries in $y$ direction: a) Quasiparticle dispersion of a single hole for the $t-J$ system obtained with the RNN (blue markers), compared to the MPS spectral function from Ref. \cite{Bohrdt2020} with the spectral weight $S$ indicated by the colormap and shown in the inset figure for $\vec{k}=(0.44\pi, 0.44\pi)$. We average the energy over the last $100$ training iterations, each with $200$ samples, with the respective error bars denoted in blue. b) Dispersion of the $t-J_z$ system obtained with the RNN, compared to the MPS spectral function. c) Relative errors $\Delta \epsilon= \frac{E_\mathrm{RNN}-E_\mathrm{DMRG}}{\vert E_\mathrm{DMRG}\vert}$ during the training for $t-J$ and $t-J_z$ systems, both with $d_h=300$. Dashed vertical lines denote the training step where the training was restarted. In the last restart the number of samples per minimization step was increased from $200$ to $600$ ($t-J$) or $1000$ ($t-J_z$).}
\label{fig:1}
\end{figure}

In the absence of doping ($\hat{n}_{\vec{i}}=1$), Eq. \eqref{eq:tJ} reduces to the XXZ model or, in the case of $J_z=J{\pm}$, the Heisenberg model. Prior studies have already utilized RNNs to simulate these spin models \cite{Roth2020,HibatAllah2022}, with the possibility of rendering the model stoquastic by making use of the Marshall sign rule
\cite{Marshall1955}. This is done by implementing the sign rule directly in the RNN architecture \cite{HibatAllah2022}, yielding a simplified optimization procedure of the wave functions' phase.

When the ground state at $\hat{n}_{\vec{i}}=1$ is doped with a single hole, the resulting mobile impurity gets dressed with a cloud of magnetic excitations. This yields the formation of a magnetic polaron, which has already been observed in ultracold atom experiments \cite{Koepsell:2019}. Its properties strongly depend on the spin background, see Fig. \ref{fig:1}a and b. Upon further doping, the strong correlations in the model make the simulation of the Fermi-Hubbard or $t-J$ models numerically challenging, despite impressive numerical advances in the past years \cite{Qin2020,Schaefer2021,xu2023,Arovas2022}: Commonly used methods all come with their specific limitations, e.g. density matrix renormalization group \cite{Schollwoeck2011,White1992} is limited by the area-law of entanglement, making it challenging to apply this methods to 2D or higher dimensions. Finally, the calculation of spectral functions or the dispersion relations $E(\vec{k})$ \cite{Bohrdt2020}, as exemplary shown in Fig. \ref{fig:1}, is of great interest for many fields in physics to reveal emergent physics of a system under investigation. In condensed matter physics, they are typically used to infer the dominating excitations in the ground state or higher energy states, e.g. upon doping the system. This information is contained in specific features of the spectra, e.g. the bandwidth of the quasiparticle dispersion $E(\vec{k})$. However, the calculation of spectra or dispersions $E(\vec{k})$ is in general computationally costly using conventional methods, e.g. density-matrix renormalization group (DMRG) simulations: The former typically involves a, in general expensive, time-evolution of the state \cite{Damme2021}, and the latter the calculation of a global operator, the momentum $\vec{k}$, which is typically very costly for matrix-product-states.  \\

The remaining part of the paper is structured as follows: In the first section, we introduce the fermionic RNN architecture and its training. Second, we apply the RNN architecture for the ground state search of the $t-$XXZ model on different lattice geometries, including 1D and 2D lattices. Furthermore, we present a method to map out the dispersion relation of the system under consideration. This method is not limited to our specific RNN quantum state representation, but applicable for any NQS architecture. Moreover, it can in principle be combined with spatial symmetries, that potentially help to improve the accuracy, and furthermore enable the analysis of low-lying excitations in a specific symmetry sector, e.g. $m_4$ rotational resonances \cite{Bohrdt2018,bohrdt2023dichotomy}. We present the results for different lattice geometries, including a triangular ladder. Finally, we address the limitations and drawbacks of our RNN ansatz, provide tests on the effects of more sophisticated training procedures, and discuss possible improvements.

\section{Architecture and training \label{sec:RNN}}

In the present paper we use a recurrent neural network (RNN) \cite{Hochreiter1997} to represent a quantum state defined on a 2D lattice with $N_{\mathrm{sites}}=N_x\cdot N_y$ positions occupied by $N_p$ particles. RNNs and similar generative architectures combined with variational energy minimization have already been applied successfully for spin systems \cite{HibatAllah2020, Roth2020,Czischek2022,Carrasquilla2019}. One of the advantages of these architectures is their autoregressive property, which allows extremely efficient independent sampling from the RNN wave function \cite{Wu2019,Goodfellow2016}, which is important for the training procedure.

In order to represent fermionic wave functions, we start from the same approach as for bosonic spin systems and use an RNN architecture consisting of $N_{\mathrm{sites}}$ (tensorized) gated recurrent units (GRUs), each one representing one site of the system. The information is passed from the first cell, corresponding to the first lattice site, to the last site in a recurrent fashion, see Fig. \ref{fig:RNN} in Appendix \ref{appendix:RNNs}.\\ 

The RNN architecture and its application to model quantum states can most easily be understood for 1D systems: At each lattice site $i$ we define $\vec{\sigma}_i$, a $N_s\times d_v$ matrix, to denote the $N_s$ local sample configurations at the respective site, and $\vec{\sigma}$ the complete configuration of system size $L$, a $N_s\times L \times d_v$ matrix, with $d_v$ the visible dimension. For the $t-J$ model, each (local) configuration consists of zeros, ones and minus ones to denote holes, spin up and spin down particles, respectively, i.e. the visible dimension is $d_v=3$. Furthermore, we define the \textit{hidden state} $\vec{h}_i$ of dimension $N_s\times d_h$ that is used to pass information from previous lattice sites through the network, with $d_h$ the hidden dimension. Given the configuration $\sigma_i$ at site $i$ and a hidden state $\vec{h}_{i-1}$, the RNN cell outputs the updated hidden state $\vec{h}_i$ as well as a conditional probability distribution and a local phase. Hereby, the hidden dimension $d_h$ determines the number of parameters of our RNN quantum state.

Since it is possible to generate $N_s\geq 1$ samples at once, by passing sets of local configurations $\vec{\sigma}_i$ through the network in parallel, we will use the notation as vectors $\vec{\sigma}_i$ and $\vec{\sigma}$ in the following, where each entry in $\vec{\sigma}$ ($\vec{\sigma}_i$) corresponds to one configuration (local configuration).\\

The RNN wave function is represented by an RNN with cells that have two output layers, one for the local phase $ \phi_{\vec{\lambda}}(\sigma_i\vert \sigma_{<i})$, and one for the local amplitude $ P_{\vec{\lambda}}(\sigma_i\vert \sigma_{<i})$ \cite{HibatAllah2020}. In total, the RNN wave function is given by
\begin{align}
\ket{\psi}_{\vec{\lambda}}=\sum_{\vec{\sigma}}\mathrm{exp}(i\phi_{\vec{\lambda}}(\sigma))\sqrt{P_{\vec{\lambda}}(\sigma)}\ket{\sigma},
    \label{eq:RNNstate}
\end{align}
where $\phi_{\vec{\lambda}}(\sigma) = \sum_i^N \phi_{\vec{\lambda}}(\sigma_i\vert \sigma_{<i})$ is the phase and $\sqrt{P_{\vec{\lambda}}(\sigma)}$ with $P_{\vec{\lambda}}(\sigma) = \Pi_i^N P_{\vec{\lambda}}(\sigma_i\vert \sigma_{<i})$ is the amplitude of the respective configuration $\sigma$.

In the present work we use the tensorized 2D version of the RNN wave function introduced above, as proposed in Ref. \cite{HibatAllah2021}, where the information encoded in the hidden states is passed in a 2D manner, see Appendix \ref{appendix:RNNs}. Furthermore, we use a variant of a gated recurrent unit (GRU) instead of a simple RNN cell, that are more successful in capturing long-term dependecies \cite{Bengio1994,Schaefer2006,Razvan2013}. \\

Our RNN ansatz uses $U(1)=U(1)_{\hat{N}}\times U(1)_{\hat{S}_z}$ symmetry, i.e. conserved total particle and total magnetization, as in Refs. \cite{Roth2020, HibatAllah2020, Morawetz2021,HibatAllah2022, Barret2022,Malyshev2023}. Further details on the RNN architecture can be found in Appendix \ref{appendix:RNNs}. Moreover, in contrast to previous RNN works on the Heisenberg model \cite{HibatAllah2020}, we do not implement any bias on the phase of the quantum state such as the Marshall sign rule \cite{Marshall1955}, in order to make our architecture applicable to any number of holes in the system.

\subsection{Minimization Procedure \label{sec:minimization}}
In order to find the ground state of the system under consideration, we use the variational Monte Carlo (VMC) minimization of the energy \cite{Becca2017,Goodfellow2016}. VMC has already been used in a wide range of machine learning applications (see e.g. Refs. \cite{Carrasquila2021,Melko2019} for an overview). In VMC, the expectation value of the energy of the RNN trial wave function,
\begin{align}
\langle E_{\vec{\lambda}}\rangle = \sum_{\vec{\sigma}}\vert \psi_{\vec{\lambda}}(\sigma)\vert^2\, E^{\mathrm{loc}}_{\vec{\lambda}} (\sigma),
\label{eq:E}
\end{align}
is minimized. Here, we have defined the local energy
\begin{align}
    E^{\mathrm{loc}}_{\vec{\lambda}} (\sigma)=\frac{\langle \sigma\vert\mathcal{H}\vert\psi_{\vec{\lambda}}\rangle }{\langle \sigma \vert\psi_{\vec{\lambda}}\rangle}\, .
    \label{eq:Eloc}
\end{align}
As shown e.g. in Refs. \cite{HibatAllah2020,Inui2021} one can use the cost function
\begin{align}
    \mathcal{C} = \sum_{\vec{\sigma}}\vert \psi_{\vec{\lambda}}(\sigma)\vert ^ 2 \underbrace{\left[  E^{\mathrm{loc}}_{\vec{\lambda}} (\sigma)-\langle E^{\mathrm{loc}}_{\vec{\lambda}}\rangle\right]}_{=:-\sqrt{N_s}\Bar{\epsilon}(\sigma)} 
    \label{eq:Cost}
\end{align}
to minimize both the local energy as well as the variance of the local energy to make the training more stable. In Eq. \eqref{eq:Cost}, we have defined $\Bar{\epsilon}(\sigma):=-\frac{1}{\sqrt{N_s}}\left[  E^{\mathrm{loc}}_{\vec{\lambda}} (\sigma)-\langle E^{\mathrm{loc}}_{\vec{\lambda}}\rangle\right]$, where $N_s$ denotes the number of samples. \\

One of the main difficulties of neural network quantum states is the optimization of Eq. \eqref{eq:Cost}, due to its typically rugged landscape with many local minima and saddle points \cite{Bukov2021}. If not stated differently, we use the Adam optimizer \cite{kingma2017adam} for the optimization of Eq. \eqref{eq:Cost}, following previous works on NQS using RNNs \cite{HibatAllah2020,Morawetz2021,Roth2020}. To improve the optimization, often stochastic reconfiguration (SR) \cite{Stokes2020,Sorella1998} is used. In this method, each parameter ${\vec{\lambda}}_k$ of the neural network is optimized individually according to 
\begin{align}
    \Bar{O}_{\vec{\sigma}k} \, \delta{\vec{\lambda}}_k=\Bar{\epsilon}(\vec{\sigma})\, ,
    \label{eq:SR}
\end{align}
with $O_{\vec{\sigma} k}= \frac{1}{\psi(\vec{\sigma})}\frac{\partial \psi(\vec{\sigma})}{\partial {\vec{\lambda}}_k}$ and $\Bar{O}_{\vec{\sigma} k} = (O_{\vec{\sigma} k} - \langle O_{\vec{\sigma} k}\rangle)/\sqrt{N_s}$. In the cases where SR is applied, we use the two recently proposed, SR variants, namely minimum-step stochastic reconfiguration (minSR) and the SR variant based on a linear algebra trick by Rende et al. \cite{rende2023simple}. Both enable the use of a large numbers of NQS parameters, see Appendix \ref{appendix:minSR}. In the minSR update, Eq. \eqref{eq:SR} is solved by
\begin{align}
    \delta {\vec{\lambda}}_k = \Bar{O}^\dagger_{k \vec{\sigma}^\prime }(T^{-1})_{ \vec{\sigma}^\prime \vec{\sigma}}\, \Bar{\epsilon}(\vec{\sigma})\,,
    \label{eq:minSR}
\end{align}
with $T=\Bar{O}\Bar{O}^\dagger$ \cite{chen2023efficient}. In the version of Rende et al.,
\begin{align}
    \delta {\vec{\lambda}}_k = X_{k\vec{\sigma}^\prime}(X^TX)^{-1}_{\vec{\sigma}^\prime \vec{\sigma}}\vec{f}_{\vec{\sigma}}\,,
    \label{eq:rende}
\end{align}
with $X=\mathrm{Concat}(\mathrm{Re}\,\Bar{O}, \mathrm{Im}\,\Bar{O} ) $ and $\vec{f}_{\vec{\sigma}}=\mathrm{Concat}(\mathrm{Re}\,\Bar{\epsilon}(\vec{\sigma}), -\mathrm{Im}\,\Bar{\epsilon}(\vec{\sigma}) )$ \cite{rende2023simple}.

\subsection{Fermionic RNN Wave Functions}

\begin{figure}
\centering
\includegraphics[width=0.45\textwidth]{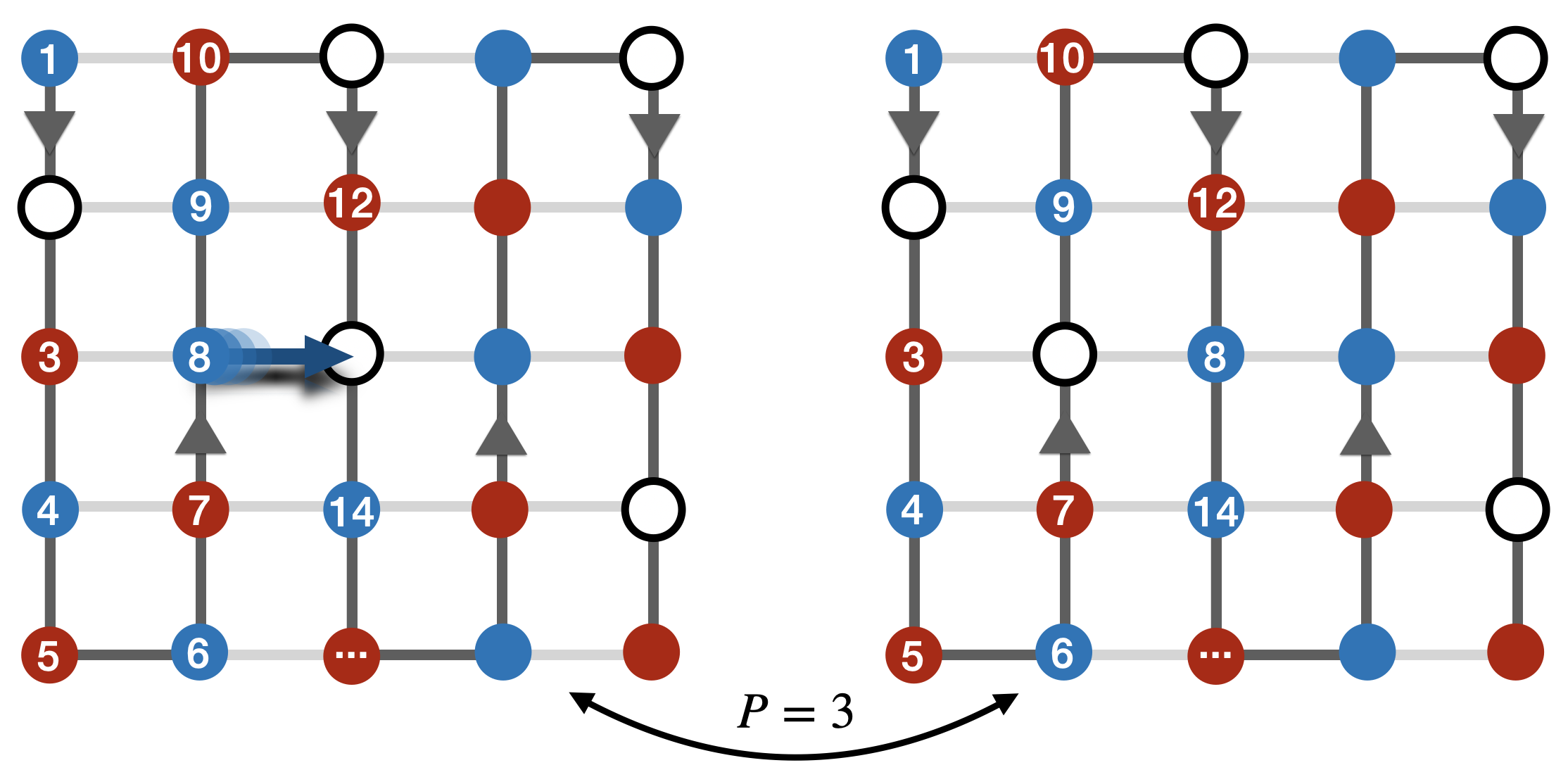}
\caption{\label{fig:SignChanges} Left: A typical configuration $\sigma$ for a $5\times5$ system with five holes and ten spin up (red) and spin down (blue) particles each. Sites are labeled in a 1D manner, as denoted by the white numbers. Right: An exemplary hopping process to the nearest neighbor in horizontal direction ends in the configuration $\sigma^\prime$ and effectively exchanges $P$ particles, here $P=3$. The respective sign of $\sigma^\prime$ relative to $\sigma$ is calculated using Eq. \eqref{eq:Eloc_antisym}.}
\end{figure}

The architecture introduced above is per se bosonic. When considering fermionic systems, we need to take the antisymmetry of the wave function into account. This antisymmetry is included during the variational Monte Carlo steps when calculating the local energy introduced in Eq. \eqref{eq:Eloc}. We can expand the local energy to
\begin{align}
    E_{loc}(\sigma)=\sum_{\vec{\sigma}^\prime}\frac{\bra{\sigma}H\ket{\sigma^\prime}\langle\sigma^\prime \vert \psi_{\vec{\lambda}} \rangle}{\langle\sigma \vert \psi_{\vec{\lambda}} \rangle}.
    \label{eq:Eloc_antisym}
\end{align}
In this sum, we multiply each term with a factor $(-1)^P$ if $\sigma^\prime$ is connected to $\sigma$ by $P$ two-particle permutations, as suggested in Ref. \cite{Inui2021}. In order to do so, we take the permutations along the sampling path into account. For the $t-$XXZ Hamiltonian under consideration we only need to consider the hopping term for calculating the antisymmetric signs. An example is shown in Fig. \ref{fig:SignChanges}.
This procedure is similar to the implementation of Jordan-Wigner strings as e.g. in Ref. \cite{Barret2022}.

\section{NQS dispersion relations \label{sec:dispersion}}
\begin{figure}
\centering
\includegraphics[width=0.48\textwidth]{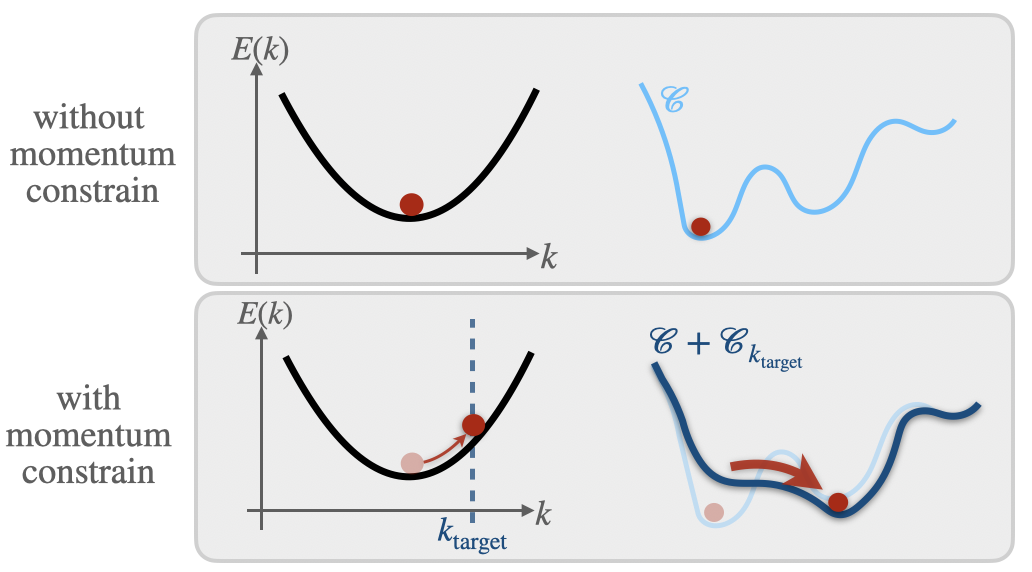}
\caption{\label{fig:Momentum} Adding the momentum constrain $\mathrm{\mathcal{C}_{k_\mathrm{target}}}$, Eq. \eqref{eq:momentumcost}, on top of the energy minimization $\mathcal{C}$, Eq. \eqref{eq:Cost}, (top right) changes the loss landscape as schematically shown on the bottom right and forces the NQS into a higher energy state with the desired momentum $k_\mathrm{target}$ (top left vs. bottom left).} 
\end{figure}

A lot of information on a physical system under investigation is contained in its dispersion relation $E(\vec{k})$, e.g. in the bandwidth (effective mass) and low-lying elementary excitations relative to the ground state, that determine the physical properties. Hence, it is of high relevance to access $E(\vec{k})$. However, its calculation is in general computationally costly \cite{Vanderstraeten2015}, since it typically requires a time-evolution of the state \cite{Damme2021}.\\

In this section, we calculate the dispersion relations $E(\vec{k})$ of $t-$XXZ models in different dimensions and on different lattice geometries using NQS. Specifically, we use the RNN wave function introduced in Sec. \ref{sec:RNN}. However, the method is applicable to any NQS architecture, in contrast to e.g. Ref. \cite{Choo2018}. It only requires the possibility to draw samples from the NQS and calculate the respective probabilities, making the calculation of $E_\mathrm{NQS}(k_x,k_y)$ computationally efficient. Furthermore, the scheme can also be combined with spatial symmetries, as discussed further in Sec. \ref{sec:symmetries}. This could help to improve the accuracy, e.g. when using a NQS with implemented translational invariance, but additional symmetries could also be used to calculate e.g. $m_4$ rotational resonances \cite{Bohrdt2018}\\

In order to calculate the dispersion relation from the NQS under consideration, we train our NQS to represent the ground state and then turn on a constrain in the loss function that forces the system to a higher energy state with the respective target momentum, see Fig. \ref{fig:Momentum}. 

The momentum $\vec{k}_\mathrm{NQS}$ of the NQS wave function is calculated from the translation operator $\hat{T}_{\vec{R}}$, which translates a state $\psi(\vec{r})$ by the respective vector $\vec{R}$, i.e. $\hat{T}_{\vec{R}} \psi(\vec{r}) = \psi(\vec{r}+\vec{R})$. Furthermore, it can be written as \cite{Shankar}
\begin{align}
    \hat{T}_{\vec{R}} = \mathrm{exp}\left(-i \vec{R}\cdot \vec{\hat{k}} \right)\,,
\end{align}
with the momentum operator $\hat{\vec{k}}$.
To determine the expectation value $\vec{k}_\mathrm{NQS}=(k_x, k_y)$ using samples $\vec{\sigma}$ drawn from the NQS wave function, we calculate the expectation value of $\hat{T}_{\vec{R}}$. For example, for a square lattice, this is done by translating all snapshots by $\vec{R}=\vec{e}_x$ and $\vec{R}=\vec{e}_y$ with $\vert \vec{e}_\mu\vert = a$ for lattice distance $a$ and $\mu=x,y$. Then, we calculate the respective NQS amplitudes of the translated states, $P_{\vec{\lambda}}(\hat{T}_{\vec{e}_{\mu}}\sigma)$, to determine the expectation value
\begin{align}
\bra{\psi_{\vec{\lambda}}}\hat{T}_{\vec{e}_\mu}\ket{\psi_{\vec{\lambda}}} = \frac{1}{N_s}\sum_{\vec{\sigma}} \frac{P_{\vec{\lambda}}(\hat{T}_{\vec{e}_\mu}\sigma)}{P_{\vec{\lambda}}(\sigma)}=\mathrm{exp}\left(-i \vec{e}_\mu\cdot \vec{k}_\mathrm{NQS} \right),
\end{align}
with the last equality due to the translational invariance of the ground state of a square lattice, which we assume to be (approximately) present for our NQS ground states, see also Appendix \ref{appendix:RNNdispersion}. Hence,
\begin{align}
\vec{k}_\mathrm{NQS}^\mu = \frac{i}{a}  \mathrm{log}\bra{\psi_{\vec{\lambda}}}\hat{T}_{\vec{e}_\mu}\ket{\psi_{\vec{\lambda}}}.
\end{align}

Using a sufficiently converged NQS ground state wave function as initial state, we train using VMC with an additional term in the loss function, 
\begin{align}
    \mathcal{C}({\vec{k}_\mathrm{target}})= \gamma(t)\sum_\mu\left(\vec{k}^\mu_\mathrm{NQS}-\vec{k}^\mu_\mathrm{target}\right)^2,
    \label{eq:momentumcost}
\end{align}
with the RNN momentum $\vec{k}_\mathrm{NQS}$ and the target momentum $\vec{k}_\mathrm{target}$. We use a prefactor $\gamma(t)= \gamma_0 \mathrm{log}_{10}(1+9(t-t_\mathrm{warmup})/\tau)$ that is turned on with typically $\tau=100,\dots, 1000$ and $\gamma_0=1,\dots,10$ and gradually lifts all areas in the loss landscape that correspond to a NQS wave function with momentum $\vec{k}_\mathrm{NQS}\neq \vec{k}_\mathrm{target}$, forcing the NQS to a higher energy state at momentum $\vec{k}_\mathrm{NQS}= \vec{k}_\mathrm{target}$, see Fig. \ref{fig:Momentum}.

For $\vec{k}_\mathrm{target}$ far away from the ground state momentum, we observe empirically that the imaginary part of $\vec{k}_\mathrm{NQS}$ can become large, on the same order as the real part, in particular if the ground state accuracy was not sufficiently high. In these cases, the RNN ends up in states that are not eigenstates of the momentum operator. In order to prevent our RNN wave function to get trapped in these states we apply an additional constrain in the loss function in these cases, penalizing large imaginary parts of the momentum, $\mathrm{Im}\, \vec{k}_\mathrm{NQS}$.

\subsubsection{$t-$XXZ model in 1D}
\begin{figure}
\centering
\includegraphics[width=0.45\textwidth]{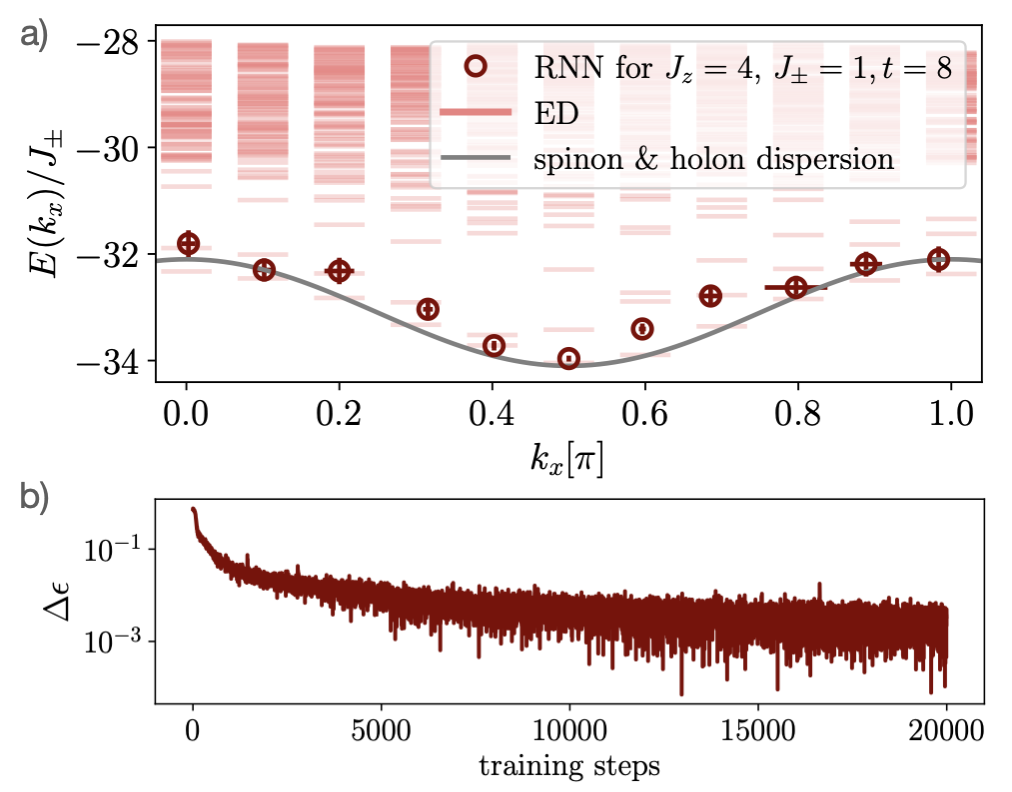}
\caption{1D $t-$XXZ system with 20 sites and $J_\pm=1$, $J_z=4$ and $t=8$: a) Quasiparticle dispersion for a single hole obtained with the RNN (red markers), compared to exact energies from ED (light red lines) and the combined spinon and holon dispersions from Eq. \eqref{eq:Ek_1D} (gray). We average the RNN energy over the last $100$ training iterations, each with $200$ samples, with the errors denoted by the errorbars. We show the exact low-energy excited states as well. b) Relative error $\Delta \epsilon= \frac{E_\mathrm{RNN}-E_\mathrm{ED}}{\vert E_\mathrm{ED}\vert}$ during the ground state training. a) and b) are obtained using a 1D RNN architecture with $d_h=100$.}
\label{fig:20x1_dispersion}
\end{figure}

In Fig. \ref{fig:20x1_dispersion}a the dispersion for an antiferromagnetic $t-$XXZ chain with $20$ sites and $J_\pm=1$, $J_z=4$ and $t=8$, obtained with a 1D RNN and exact diagonalization (ED) is shown. The relative error on the ground state energy at $k_x=0.5\pi$, obtained during a training with $20000$ iterations, is shown in Fig. \ref{fig:20x1_dispersion}b. The energies away from the ground state at $k_x=0.5\pi$, see Fig. \ref{fig:20x1_dispersion}a, are in relatively good agreement with the exact values from ED. However, at some values of $k_x\neq 0.5 \pi$ it can be seen that the RNN is trapped in local minima close to the ground state. Overall, the RNN succeeds in capturing physical properties like the bandwidth very accurately, revealing the underlying physical excitations: 

For the system under consideration, the bandwidth and the shape of the dispersion in Fig. \ref{fig:20x1_dispersion}a is a result of spin-charge separation in 1D systems. Spin-charge separation denotes the fact that the motion of a hole in such an AFM spin chain with coupling $J_\pm, J_z\ll t$ can be approximated by an almost free hole that is only weakly coupled to the spin chain. Hence, the dispersion in Fig. \ref{fig:20x1_dispersion} can be approximated by two separate dispersions; i.e. holon and spinon dispersions. Hereby, the holon is the charge excitation, associated with energy scales $t$, and the spinon is the spin excitation associated with energy $J_\perp, J_z$. In Ref. \cite{Bohrdt2018} it is shown that the combined dispersion is 
\begin{align}
    E(k_x)=-2t \,\mathrm{cos}(k_h)+J_\pm \, \mathrm{cos}\left(2(k_x-k_h)\right)+J_\pm+J_z,
    \label{eq:Ek_1D}
\end{align}
where $k_h$ is the momentum of the holon and $k_x=k_h+k_s$ is the combined momentum of holon and spinon. Eq. \eqref{eq:Ek_1D} is denoted by the gray line in Fig. \ref{fig:20x1_dispersion}. Again, the agreement with the RNN is relatively good.

\subsubsection{$t-J$ model on a square lattice}

\begin{figure}[t]
\centering
\includegraphics[width=0.5\textwidth]{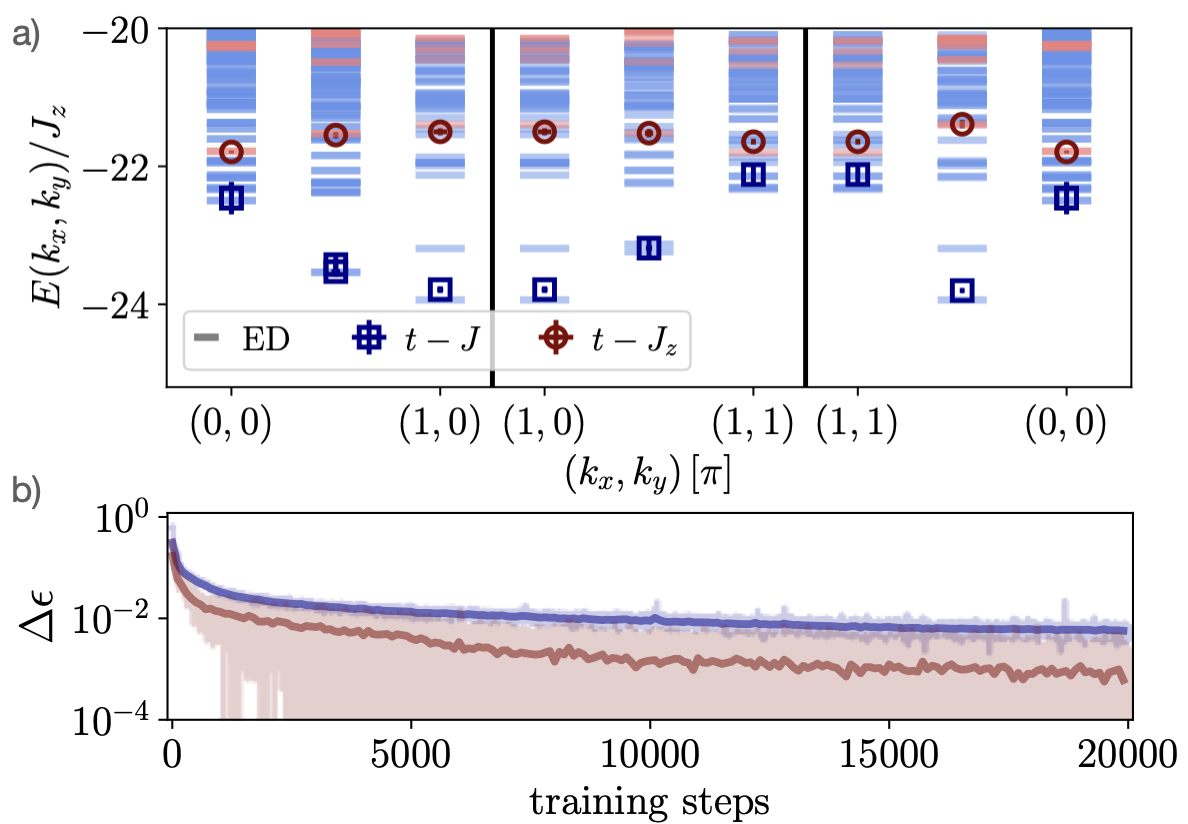}
\caption{$t-J$ (blue) and $t-J_z$ (red) square lattice with $4\times 4$ sites, $t/J=3$ and periodic boundaries: a) Quasiparticle dispersion for a single hole obtained with the RNN (blue and red markers), compared to the exact energies from ED (lines). We average the energy over the last $100$ training iterations, each with $200$ samples, with the respective error bars shown in blue and red. We show the exact low-energy excited states as well. b) Relative error $\Delta \epsilon$ during the ground state training for $t-J$ (light blue) and $t-J_z$ (light red) square lattice ground states, with $d_h=100$ and minSR ($t-J$) and $d_h=70$ and Adam ($t-J_z$). Thick lines are averages over $100$ training iterations to guide the eye.}
\label{fig:4x4dispersion}
\end{figure}

Due to the layered structure of high-T$_c$ superconductors like cuprates \cite{Keimer:2015} or nickelates \cite{Li2019,Sun2021}, the physics of $t-J$ systems upon doping is particularly interesting in 2D.
In Figs. \ref{fig:1} and \ref{fig:4x4dispersion}, the Quasiparticle dispersion for a single hole on $10\times 4$ and $4\times 4$ $t-J$ and $t-J_z$ lattices are presented. In both cases, Figs. \ref{fig:1}b and \ref{fig:4x4dispersion}b show that the ground state convergence is better for the $t-J_z$ model with relative errors on the order of $\Delta \epsilon\approx10^{-3}$ for both system sizes, yielding a good agreement with the reference energies from DMRG ($10\times 4$ system) and ED ($4\times 4$ system) for all considered energies $E(k_x,k_y)$ away from the ground state. With a relative error of $\Delta \epsilon\approx10^{-2}$, the error of the $t-J$ ground states is above the $t-J_z$ systems, which is also reflected in the accuracy of the dispersion $E_\mathrm{RNN}(k_x,k_y)$ in Figs. \ref{fig:1}a and \ref{fig:4x4dispersion}a. 

In contrast to the previous section, there is no spin-charge separation in the strict sense in two dimensional systems. In the case $t \gg J_\pm=J_z=: J$ that we consider here ($t/J=3$), the mobile dopant can be described by fractionalized spinons and chargons that are confined by a string-like potential that arises due to the spin background distortion when the dopant moves through the system \cite{Beran1996,Grusdt2018}. Based on this idea, Laughlin \cite{Laughlin1997} drew the analogy with the 1D Fermi-Hubbard or $t-J$ systems and suggested that the dispersion in the respective 2D systems can be interpreted in terms of pointlike partons, spinons and chargons, that interact with each other. This \textit{parton picture} explains that the quasiparticle dispersion for a single hole is dominated the spinon with a bandwidth on the order of $J_{\pm}$, with corrections by the chargon on energy scales of $t$ \cite{Bohrdt2020}. This mechanism also provides the explanation for the flat dispersion for the $t-J_z$ model in contrast to the $t-J$ model, as captued by the RNN, see Figs. \ref{fig:1} and \ref{fig:4x4dispersion}. Despite the small deviations from the dispersions calculated with ED or DMRG, our RNN architecture, succeeds in capturing the respective bandwidths of $t-J_z$ and $t-J$ models very accurately, allowing to gain valuable insights on the spinon and chargon physics from the RNN dispersions. Furthermore, the fact that node ($\pi/2,\pi/2$) and antinode ($\pi,0$) are degenerate in the $4x4$ system is correctly reproduced.\\

Lastly, we would like to mention that there is a small region of suppressed spectral weight near $(\pi,\pi)$ in the DMRG results of the $t-J$ system \cite{Bohrdt2018}. This suppression yields difficulties for our RNN scheme that are further discussed in Appendix \ref{appendix:RNNdispersion}.

\subsubsection{$t-J$ model on a triangular lattice \label{sec:triangular}}
\begin{figure}
\centering
\includegraphics[width=0.45\textwidth]{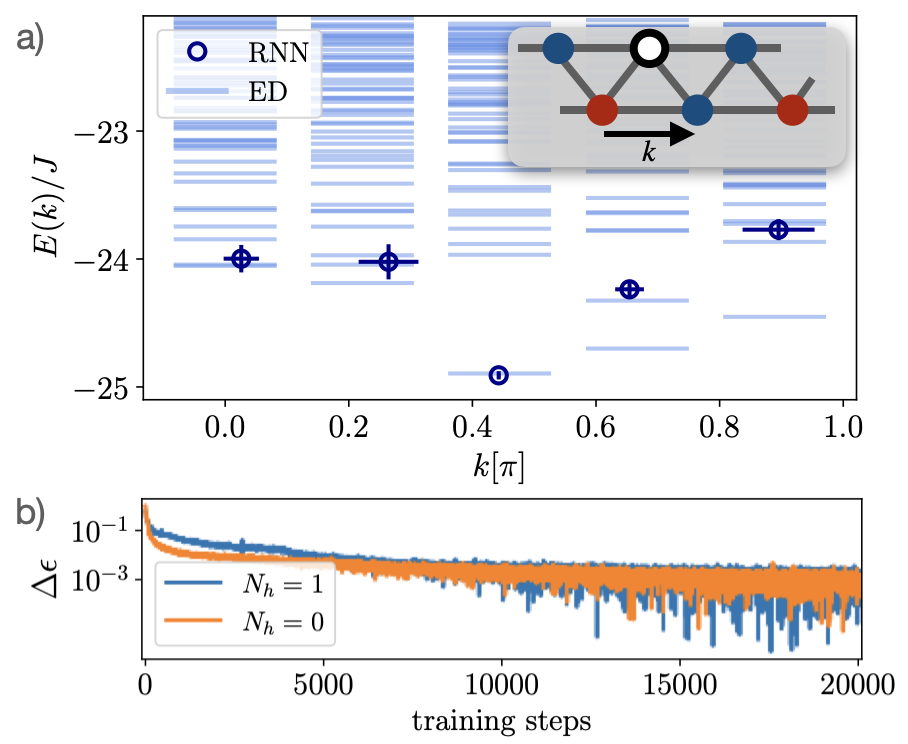}
\caption{$t-J$ model on a triangular lattice with $9\times 2$ sites, $t/J=3$ and periodic boundaries along $x$ direction: a) Quasiparticle dispersion for a single hole obtained with the RNN (blue markers), compared to the exact energies from ED (light blue lines). We average the energy over the last $100$ training iterations, each with $200$ samples, with the error denoted by the blue errorbars. We show the exact low-energy excited states as well. b) Relative error $\Delta \epsilon$ during the ground state training without doping (orange) and with one hole (blue).}
\label{fig:2x9_triangular}
\end{figure}

On triangular lattices, the physical phenomena that are observed are distinctly different from the physics of bipartite lattices, due to the notion of frustration and the absence of particle-hole symmetry in non-bipartite lattices, among them e.g. kinetic frustration \cite{Haerter2005,schlömer2023kinetictomagnetic}. In particular, the underlying constituents upon doping the triangular ladder are not known \cite{schlömer2023kinetictomagnetic}, making the triangular lattice an intriguing system to study. Recent advancements have shown that these lattices can also be studied experimentally using optical triangular lattices \cite{Struck2011,Tang:2020,Xu:2023} and solid state platforms based on Moir\'e heterostructures \cite{Yamamoto_2020,Wu2018,Davydova_2023}.\\

Triangular spin systems have already been studied using RNNs \cite{HibatAllah2022}.
Here, we consider a triangular $t-J$ ladder with length $L_x=9$, with the quasiparticle dispersion for a single hole and the learning curves with and without doping shown in Fig. \ref{fig:2x9_triangular}.

As suggested in Ref. \cite{HibatAllah2022}, we use variational annealing for the training for the triangular lattice, that was shown to improve the performance for frustrated systems like the triangular Heisenberg model \cite{HibatAllah2022}. The idea of annealing is to avoid getting stuck in local minima by including an artificial temperature $T$ in the learning process. In order to do so, the variational free energy of the model,
\begin{align}
    F_{\vec{\lambda}} = \langle H_{\vec{\lambda}} \rangle -T(n_\mathrm{step})\cdot  S\,
\end{align}
instead of the energy \eqref{eq:E} is minimized. Here, the averaged Hamiltonian $\langle H_{\vec{\lambda}} \rangle$ is given by $\langle H_{\vec{\lambda}} \rangle=\sum_{\vec{\sigma}} \vert \psi_{\vec{\lambda}}(\sigma)\vert^2 H_{\vec{\lambda}} (\sigma)$. Furthermore, $S$ denotes the Shannon entropy
\begin{align}
    S=-\sum_{\vec{\sigma}} \vert \psi_{\vec{\lambda}}(\sigma)\vert^2 \mathrm{log}\left[\vert \psi_{\vec{\lambda}}(\sigma)\vert^2 \right]\, .
\end{align}
The minimization procedure that we use starts with a warmup phase with a constant temperature $T_0$, before decreasing the temperature $T(t)=T_0(1- (t-t_\mathrm{warmup})/\tau)$ linearly with the minimization steps $t$ with $\tau=10000$ and $t_\mathrm{final}=40000$ training steps.\\

In Fig. \ref{fig:2x9_triangular}b it can be seen that this procedure yields relatively good results for the ground states, with errors of $\Delta \epsilon\approx 0.001$ for both $N_h=0$ and $N_h=1$. For the dispersion shown in Fig. \ref{fig:2x9_triangular}a, we consider the momentum $k$ defined along the ladder, as shown in the inset figure. When enforcing $k\neq 0.444\pi$ away from the ground state, the exact energy gaps from ED to the first excited states strongly decrease and the the RNN gets trapped in these states in most cases, in particular for $k> 0.444\pi$. Furthermore, the errorbars of the enforced momenta are much higher compared to the other lattice geometries that were studied in Figs. \ref{fig:1}, \ref{fig:20x1_dispersion} and \ref{fig:4x4dispersion}, suggesting that the RNN states partly break the translation invariance, and hence challenge the momentum optimization scheme.

In this section, we discuss the capability of our bosonic and fermionic RNN ansätze presented in Sec. \ref{sec:RNN} to learn and represent the ground states of the $t-$XXZ model. For our analysis, we focus on $t-J$ and $t-J_z$ models on a $4\times 4$ square lattice. 

Figs. \ref{fig:tJz_overview} and \ref{fig:tJ_overview} show the relative error for the ground state energies of $t-J_z$ and $t-J$ models obtained with our RNN ansatz upon doping the half-filled system with $N_h$ holes. Starting from $N_h=0$ in the $t-J_z$ model, the accuracy of the respective Ising ground state is very high in both cases with relative errors $\Delta \epsilon= \frac{E_\mathrm{RNN}-E_\mathrm{ED}}{\vert E_\mathrm{ED}\vert}$ below the numerical precision. The $t-J$ model, reducing to the Heisenberg model at $N_h=0$, features spin-flip terms besides the Ising interactions, making the ground state search more difficult. Our RNN reaches a ground state energy error $\Delta \epsilon \approx 10^{-4} $ after $20000$ training steps. For both models, the phase and amplitude distributions shown in Figs. \ref{fig:tJz_overview}b and \ref{fig:tJ_overview}b are relatively simple with a low variance for the logarithmic amplitude and only two values for the phase, $0$ and $\pi$. In particular, the Ising state for the $N_h=0$ case of the $t-J_z$ model, features basically only two N\'eel states with non-zero amplitude (i.e. approx. zero log-amplitudes), shown in Fig \ref{fig:tJz_overview}b on the very left.  Note that when comparing to the literature of ground state representations using RNNs for the Heisenberg model \cite{HibatAllah2020,Roth2020}, the optimization problem in our setup is more challenging due to the following reasons: $(i)$ The RNN that we use has a local Hilbert space dimension of three states instead of two, allowing for all values of $N_h$ in principle. $(ii)$ Our RNN learns the sign structure without any bias, i.e. we do not implement the Marshall sign rule already in the RNN, which would only work for $N_h=0$. $(iii)$ We do not include the knowledge of spatial symmetries yet, which will be done later in Sec. \ref{sec:symmetries}.\\

\FloatBarrier
\section{Performance of the RNN ansatz}
\begin{figure}[t]
\centering
\includegraphics[width=0.45\textwidth]{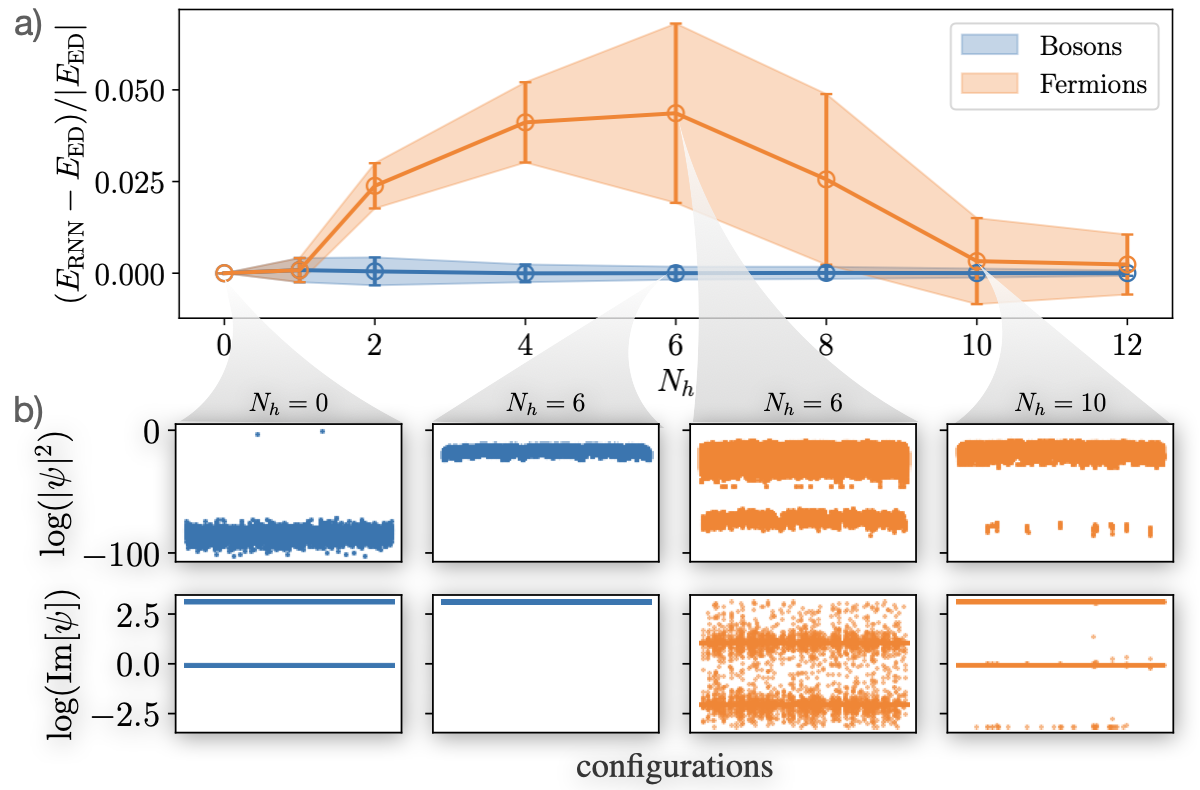}
\caption{RNN representation for ground states of the bosonic and fermionic $t-J_z$ model with $t/J_z=3$, $0\leq N_h \leq 12$ for a $4\times 4$ square lattice with open boundaries. a) Relative error for bosons (blue) and fermions (orange). b) Logarithmic amplitude and phase distributions from ED for exemplary bosonic (blue) and fermionic (orange) hole numbers. On the very left, the two states $\sigma_\text{N\'eel}$ with $\mathrm{log}\vert \psi (\sigma_\text{N\'eel})\vert^2=0$ are the N\'eel states. We use a hidden dimension of $h_d=100$.  }
\label{fig:tJz_overview}
\end{figure}

Upon doping, the relative error of the ground states without antisymmetrization of the RNN wave function for the $t-J_z$ model in Fig. \ref{fig:tJz_overview} is below $\Delta \epsilon \ll 5 \cdot 10^{-4} $ for all considered hole dopings $1\leq N_h\leq 12$. As exemplary shown for the bosonic $N_h=6$ case in Fig. \ref{fig:tJz_overview}b in blue, the true ground state from exact diagonalization does not have a phase structure in this case and the logarithmic amplitudes are very similar. When including the antisymmetry for the fermionic wave functions, the variance of both phase and amplitude distributions increases, from $\sigma_{N_h=6}^\mathrm{b}(\mathrm{log}\vert\psi\vert^2)=2.23$ to $\sigma_{N_h=6}^\mathrm{f}(\mathrm{log}\vert\psi\vert^2)=19.00$, and $\sigma_{N_h=6}^\mathrm{b}(\mathrm{Im}\psi)=0$ to $\sigma_{N_h=6}^\mathrm{f}(\mathrm{Im}\psi)=2.47$, which can be seen from bare eye when comparing the bosonic and fermionic ED distributions in Fig. \ref{fig:tJz_overview}b. This complicates the ground state search and the ground state error increases significantly between $2\leq N_h \leq 9$ for the fermionic $t-J_z$ model. At $N_h=10$, when only four particles remain in the system and probably a Fermi-liquid regime is entered, the error decreases again  to $\Delta \epsilon < 1\%$ in the fermionic case, coinciding with a lower variance of the exact log-probabilities than for $N_h=6$, $\sigma_{N_h=6}^\mathrm{b}(\mathrm{log}\vert\psi\vert^2)=9.48$. \\

\begin{figure}[t]
\centering
\includegraphics[width=0.45\textwidth]{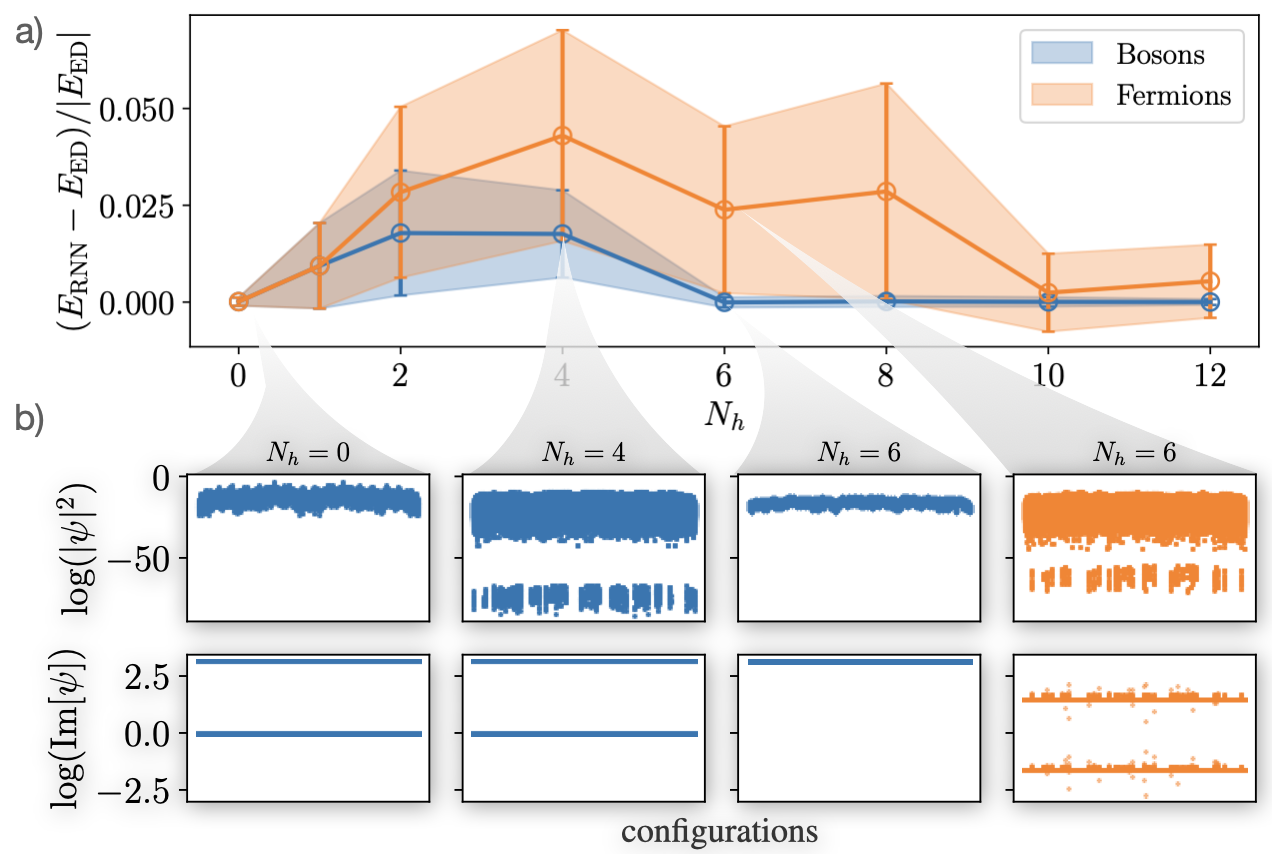}
\caption{RNN representation for ground states of the bosonic and fermionic $t-J$ model with $t/J=3$, $0\leq N_h \leq 12$ for a $4\times4$ square lattice with open boundaries. a) Relative error for bosons (blue) and fermions (orange). b) Logarithmic amplitude and phase distributions from ED for exemplary bosonic (blue) and fermionic (orange) hole numbers.  We use a hidden dimension of $h_d=100$.}
\label{fig:tJ_overview}
\end{figure}

The exact log-amplitude and phase distributions from ED for $N_h>0$ of the $t-J$ model are typically more complicated than for the $t-J_z$ model. For example, for $N_h=4$, the variance of the exact amplitudes becomes very large, $\sigma_{N_h=6}^\mathrm{b}(\mathrm{log}\vert\psi\vert^2)=15.91$, see Fig. \ref{fig:tJ_overview}b. This yields larger ground state energy errors than for the $t-J_z$ model, and is further complicated when including the antisymmetry in the fermionic case. Again, we make the observation that for larger hole dopings, $N_h\geq 6$ for bosons and $N_h\geq 10$ for fermions, the distributions for phase and amplitude become less complicated than in the low to intermediate doping regime, yielding a higher accuracy of the RNN wave function with errors $\Delta \epsilon\leq 10^{-4}$ for bosons and $\Delta \epsilon\leq 10^{-2}$ for fermions in the respective doping regimes.\\

Our results show that in the low doping regime of the $t-J$ model, both fermionic systems and bosonic systems are difficult to learn, see Fig. \ref{fig:tJ_overview}. This suggests that not only the fermionic sign structure is challenging, but also the motion of bosonic holes in the AFM Heisenberg background. When these holes move through the system, the spin background is affected, giving rise to an effective $J_1-J_2$ spin model with nearest and next-nearest spin exchange interactions and is hence more difficult to learn \cite{schlömer2022quantifying}. For the $t-J_z$ model, we observe that, probably due to the lack of spin dynamics resulting from the absence of spin-flip terms, the relative errors are comparably low in the bosonic case.

 Furthermore, for all states with high $\mathrm{log}\vert\psi\vert^2$ variance, there are several configurations $\vec{\sigma}$ with a large negative log-amplitude, i.e. $\vert\psi(\vec{\sigma})\vert^2\approx 0$. This makes an accurate determination of expectation values extremely costly and can affect the training process. For example, in Ref. \cite{sinibaldi2023unbiasing} it was shown that this yields higher variances for the gradients determined by stochastic reconfiguration.\\

Given these relatively high errors on the ground state energies in some cases, we test potential bottlenecks of our approach in the following, namely: $(i)$ Difficulties in learning either the phase or the amplitude, by considering the partial learning problems separately. $(ii)$ The optimization procedure. $(iii)$ The optimization landscape. $(iv)$ The expressivity of the RNN ansatz, compared to the complexity of the learning problem.

\subsubsection{The partial learning problem}
One potential bottleneck of our approach is the way the RNN wave function is split into amplitude and phase. In order to test if there are problems with the optimization of the phase or amplitude alone, we consider their learning problems separately as suggested e.g. in Refs. \cite{wang2023,Bukov2021}. 
\begin{enumerate}
    \item Phase training: We sample from the exact ground state distribution $\vert \psi \vert^2$, calculated with ED, and optimize only the phase.
    \item Amplitude training: Given the correct phase distribution from ED, we optimize only the logarithmic amplitude to check if the ground-state probability amplitudes can be learned.
\end{enumerate}

\begin{figure}[t]
\centering
\includegraphics[width=0.48\textwidth]{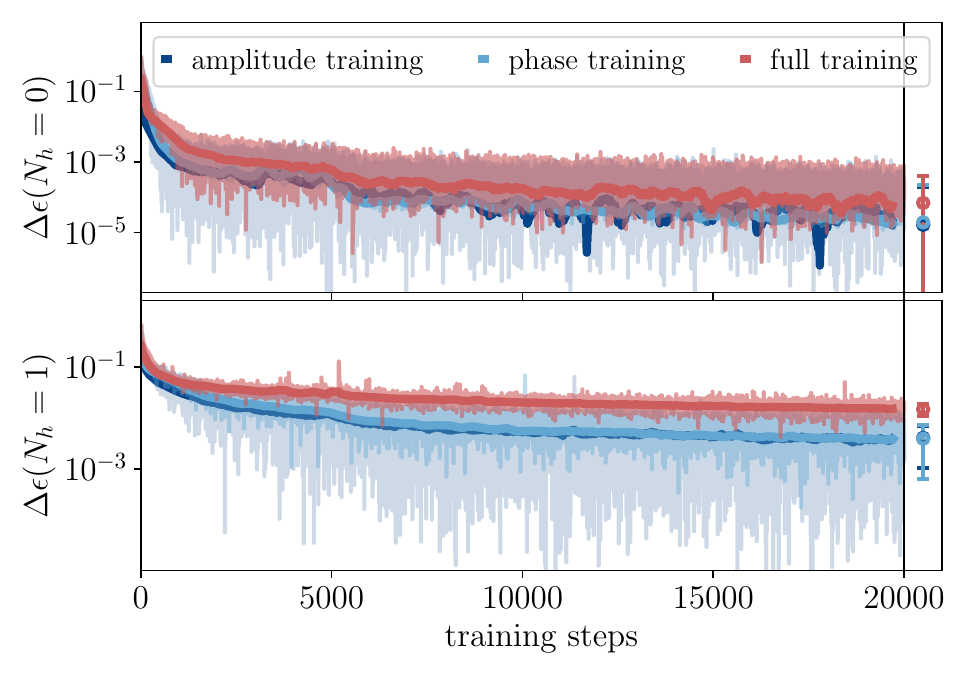}
\caption{Partial training, i.e. separate amplitude (dark blue) and phase (light blue) training, for ground states of the $t-J$ model on a $4\times 4$ square lattice with $t/J=3$, open boundaries and $N_h=0$ (top) and $N_h=1$ (bottom), compared to the full training in red.  We use a hidden dimension of $h_d=70$.}
\label{fig:partialtraining}
\end{figure}

Fig. \ref{fig:partialtraining} shows the results of amplitude and phase trainings (dark and light blue), compared to the full training of both amplitude and phase (red). For all considered systems, the results of the partial trainings are closer to the exact ground state, e.g. for open boundaries and $N_h=1$, the relative error is decreased from $\Delta\epsilon = 0.0147(37)$ to $\Delta\epsilon =0.0040(30)$ for the amplitude training and $\Delta\epsilon =0.0039(33)$ for the phase training. However, for all considered cases we observe the same problem as in the full training: the RNN gets stuck in a plateau that survives up to $20000$ training steps. Although the relative error of the plateau decreases when considering the partial learning problems, the improvement is surprisingly low given the amount of information that is added to the training. Furthermore, whether the amplitude or phase training is more problematic remains unclear. Even for the phase training, for which the training samples are generated from the exact distribution $\vert \psi\vert^2$ calculated with ED, the improvement is not significantly larger than for the amplitude training. This is in agreement with the results of Bukov et al. \cite{Bukov2021}.

\subsubsection{Comparison of optimizers}

\begin{figure}[t]
\centering
\includegraphics[width=0.49\textwidth]{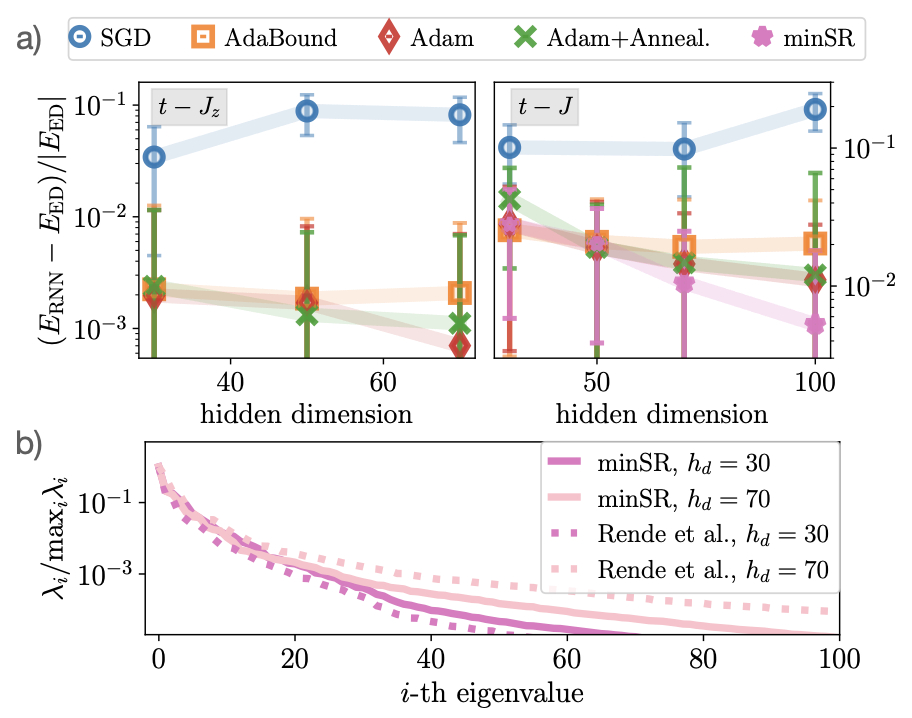}
\caption{a) Testing different optimizers: Optimization results for the $t-J_z$ model (left) and the $t-J$ model (right) on a $4\times 4$ square lattice with $t/J_z=3$, both for $N_h=1$ and periodic boundaries, using SGD, AdaBound, Adam, Adam+Annealing and minSR, and 200 samples (1000 samples for minSR) in each VMC step. b) Eigenvalues of the $T$-matrix (minSR algorithm \cite{chen2023efficient}, solid lines) and of the $X^TX$ matrix (SR variant of Rende et al. \cite{rende2023simple}, dotted lines) before the training, for the $4\times4$ $t-J$ system with one hole and open boundaries and $h_d=30,70$, using $1000$ samples, }
\label{fig:optimizers}
\end{figure}

As a next test, we compare the optimization results of different optimizers in Fig. \ref{fig:optimizers}a, namely Stochastic gradient descent (SGD), adaptive methods like AdaBound \cite{Luo2019AdaBound} and  Adam \cite{kingma2017adam}, and more advanced methods such as Adam+Annealing \cite{HibatAllah2022} and the recently developed variant of stochastic reconfiguration (SR), minimum-step SR (minSR) \cite{chen2023efficient}. We show the optimization results for the $t-J_z$ model on the left and the $t-J$ model on the right, both for $N_h=1$. \\

Typically, Adam is used for RNN wave function optimization \cite{HibatAllah2020,HibatAllah2021,HibatAllah2022,Roth2020}, adapting the learning rate in each VMC update. For $200$ samples used in each optimization step, Adam yields relative errors on the order of  $\Delta \epsilon \approx10^{-3} $ for the $t-J_z$ model and $\Delta \epsilon \approx10^{-2} $ for the $t-J$ model. AdaBound, employing dynamic bounds on learning rates, yielding a gradual transition from Adam to SGD during the training, has similar results. \\

Another modification of the Adam training is the use of variational annealing as introduced in Sec. \ref{sec:triangular}, shown to improve the performance for frustrated systems \cite{HibatAllah2022}.
The minimization procedure that we use starts with a warmup phase with a constant temperature $T_0=1$, before decreasing the temperature $T(t)=T_0(1 - (t-t_\mathrm{warmup})/\tau)$ linearly with the minimization steps $t$. Typically, we use $\tau=5000$ and stop the training after $t_\mathrm{final}=20000$ training iterations, but tests up to $\tau=20000$ and $t_\mathrm{final}=40000$ did not yield any improvements. Fig. \ref{fig:optimizers}a shows that for the square lattice, the use of annealing does not bring any advantage within the errorbars.\\

Lastly, we apply minSR, a recently developed variant of SR \cite{chen2023efficient}, as introduced in Sec. \ref{sec:minimization}. For a stable training, we ensure non-exploding gradients by adding a diagonal offset $\delta(t)$ to the diagonals of the $T$-matrix, with $\delta(t)$ exponentially decaying from $1$ to $10^{-10}$. After determining the gradients using Eq. \eqref{eq:minSR}, we apply the Adam update rule, which we empirically find to perform better than the GD update. Moreover, since it is crucial to use enough samples for a sufficiently good approximation of the gradients in SR, typically more samples than for the other optimization routines are needed. Here, we use $1000$ samples in each minSR update and find that the results on the one-hole $t-J$ ground state errors improve below the values obtained with Adam, see Fig. \ref{fig:optimizers}a on the right. However, we show in Appendix \ref{appendix:minSR} that a comparison with Adam using the same number of samples does not lead to a conclusive result which optimization routine is better, similar to the SR results in Ref. \cite{Bukov2021}. 

The reason behind this can be understood when considering the spectrum of the $T$-matrix of the minSR algorithm: Similar to the results of Ref. \cite{Donatella2023} for the $S$-matrix of the SR algorithm, Fig. \ref{fig:optimizers}b shows that the eigenvalues of $T$, $\lambda_i$, decrease extremely rapidly, in particular at the beginning of the training, indicating a very flat optimization landscape. This is a typical problem of autoregressive architectures \cite{Donatella2023} and causes uncontrolled, high values of $T^{-1}$ and consequently also of the gradients $\delta \theta$, see Eq. \eqref{eq:minSR}. Furthermore, the shape of the spectrum does not have any feature that indicates that the spectrum could be cut off at a specific eigenvalue, making a regularization very difficult. Hence, the diagonal offset $\delta(t)$ must be chosen relatively large, yielding parameter updates that are very similar to the plain vanilla Adam optimization as long as $\delta(t)$ is larger than many of the $T$-eigenvalues. The spectrum of the $(X^TX)$ matrix of the SR variant by Rende et al. \cite{rende2023simple}, see Eq. \eqref{eq:rende}, exhibits the same problem.\\

When comparing the results for different hidden dimensions, e.g. for minSR in Fig. \ref{fig:optimizers}a (right), it may suggest that a hidden dimension $h_d>100$ could in principle improve the results further. However, we will show in Sec. \ref{sec:params} that for such a large number of parameters, it is even possible, by restricting to a fixed number of holes and hence reducing the Hilbert space dimension to $\ll 3^{N_\mathrm{sites}}$, to encode the wave function using exact methods.

\subsubsection{Spatial symmetries \label{sec:symmetries}}

\begin{figure}[t]
\centering
\includegraphics[width=0.42\textwidth]{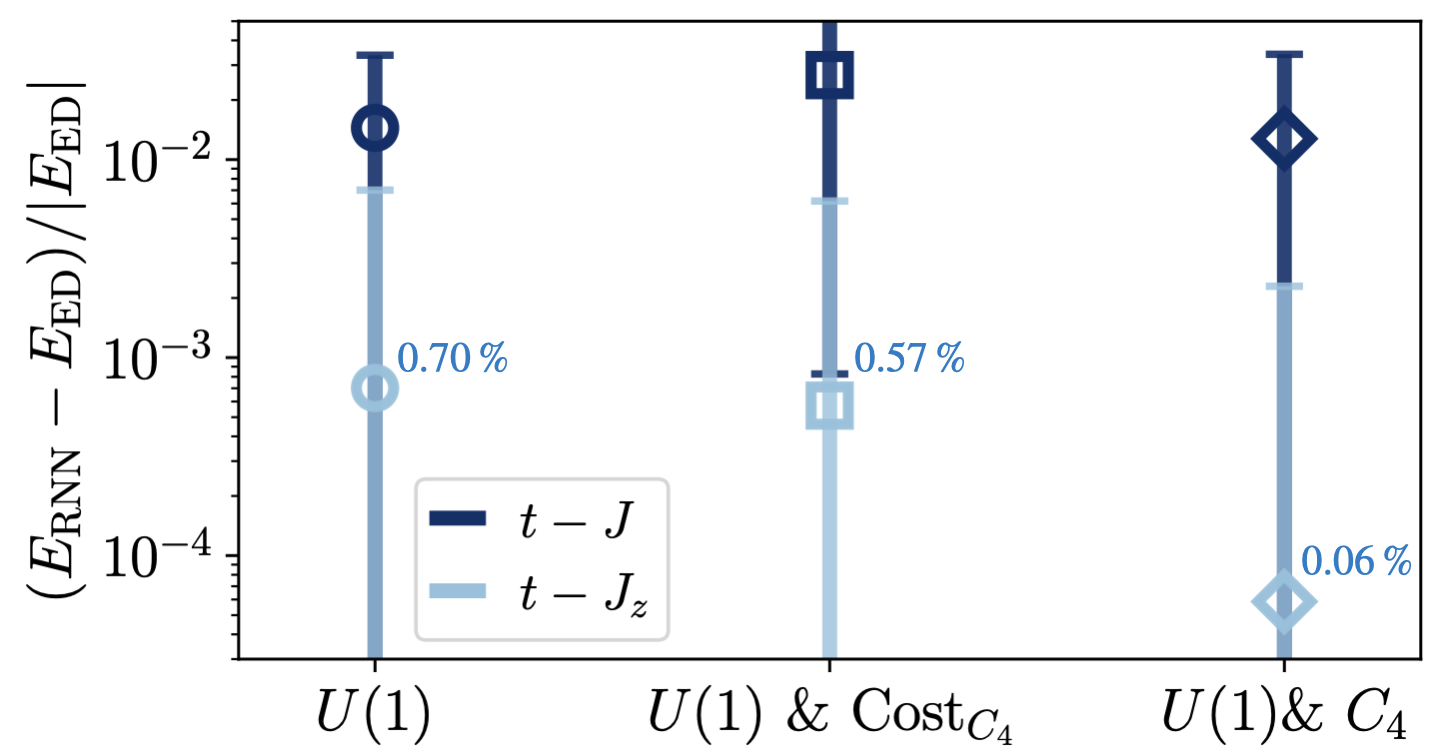}
\caption{Relative error for $t-J$ (dark blue) and $t-J_z$ (light blue) models on a $4\times 4$ square lattice with one hole, $t/J_z=3$ and periodic boundaries, for RNNs with implemented $U(1)=U(1)_{\hat{N}}\times U(1)_{\hat{S}_z}$ symmetry, $U(1)$ and $C_4$ symmetry, implemented via the cost function and the RNN ansatz. We use a hidden dimension of $h_d=70$. For the $t-J_z$ model, we provide the relative errors as numbers in light blue. }
\label{fig:symmetry}
\end{figure}

The RNN ansatz we use has implemented $U(1)=U(1)_{\hat{N}}\times U(1)_{\hat{S}_z}$ symmetry, i.e. conserved total particle and total magnetization \cite{HibatAllah2020,Barret2022}. This is done by calculating the current particle number $N_p(i)$ (magnetization $S_z(i)$) after the $i$-th RNN cell during the sampling process and assigning a zero conditional probability if $N_p(i)=N_\mathrm{target} $ ($S_z(i)=S_{z,\mathrm{target}} $) for all sites $j>i$ that are considered afterwards, see Appendix \ref{appendix:symmetry}. As a next test, we employ additional spatial symmetries: For a symmetry operation $\mathcal{T}$ according to the lattice symmetry, we know that 
\begin{align}
    \vert \psi(\sigma)\vert^2 = \vert \psi(\mathcal{T}\sigma)\vert^2
\end{align}
for the exact ground state. For rotational $C_4$ symmetry of the square lattice, we employ this constrain 
$(i)$ in the training, by implementing it in the cost function, or $(ii)$ in the RNN ansatz as in Ref. \cite{HibatAllah2020}.

The constrain in the cost function that we use in $(i)$ is calculated by rotating all samples drawn from $\vert \psi_\mathrm{\vec{\lambda}}\vert^2$ according to $C_4$ in each VMC step, calculating $p_\mathrm{\vec{\lambda}}(\mathcal{T}_i\sigma)= \vert \psi_\mathrm{\vec{\lambda}}(\mathcal{T}_i\sigma)\vert^2$ for all $\{\mathcal{T}_i\}_i$ and adding the squared difference $\gamma(t) \sum_{\vec{\sigma}}\left( \vert \psi_\mathrm{\vec{\lambda}}(\sigma)\vert^2-\vert \psi_\mathrm{\vec{\lambda}}(\mathcal{T}_i\sigma)\vert^2\right)^2$ with a prefactor $\gamma(t)= \gamma_0 \mathrm{log}_{10}(1+9(t-t_\mathrm{warmup})/\tau)$ to the cost function. Typically, we use long decay times on the order of $\tau=5000$ steps.

For $(ii)$, we assign 
\begin{align}
    p_\mathrm{\vec{\lambda}}(\sigma)=\frac{1}{\vert \{\mathcal{T}_i\}_i\vert }\sum_{ \mathcal{T}=1, \{\mathcal{T}_i\}_i }\vert \psi_\mathrm{\vec{\lambda}}(\mathcal{T}\sigma)\vert^2
\end{align}
for all operations $\mathcal{T}_i$ in the symmetry group, similar to Ref. \cite{HibatAllah2020}.\\

The optimization results using $(i)$ and $(ii)$ are shown in Fig. \ref{fig:symmetry} for the $t-J$ and $t-J_z$ model on a $4\times 4$ square lattice. It can be seen that constraining the RNN wave function directly via $(ii)$ is more succesful than via the cost function $(i)$: Using $(ii)$, we get an order of magnitude lower relative errors compared to the results without spatial symmetries for the $t-J_z$ model. This possibly results from the fact that the additional constrain on the symmetry leads to barriers in the loss landscape in the regions where the symmetry is violated. Even when increasing the symmetry constrain gradually during the training, as described above, these barriers can prevent getting close to the minimum.

The $t-J$ model results do not improve significantly for both symmetry implementations $(i)$ and $(ii)$, with an error on the order of $\Delta \epsilon\approx 10^{-2}$ with and without spatial symmetries. Hence, we conclude that applying symmetries does only help to improve the accuracy if the ground state can already be learned sufficiently well, as for the $t-J_z$ model. 

For systems with sufficiently high convergence, also rotational symmetries like $s$, $p$ or $d$-wave symmetries could be enforced to probe the competition between the ground state energies in the respective symmetry sectors \cite{Leung2002}, which is highly relevant for the study of high-T$_c$ superconductivity. In addition, also low-energy excited states for these symmetry sectors could be calculated by making use of the dispersion scheme from Sec. \ref{sec:dispersion}, e.g. $m_4$ rotational spectra \cite{bohrdt2023dichotomy}.

\subsubsection{Complexity of the learning problem \label{sec:params}}

\begin{figure}[t]
\centering
\includegraphics[width=0.45\textwidth]{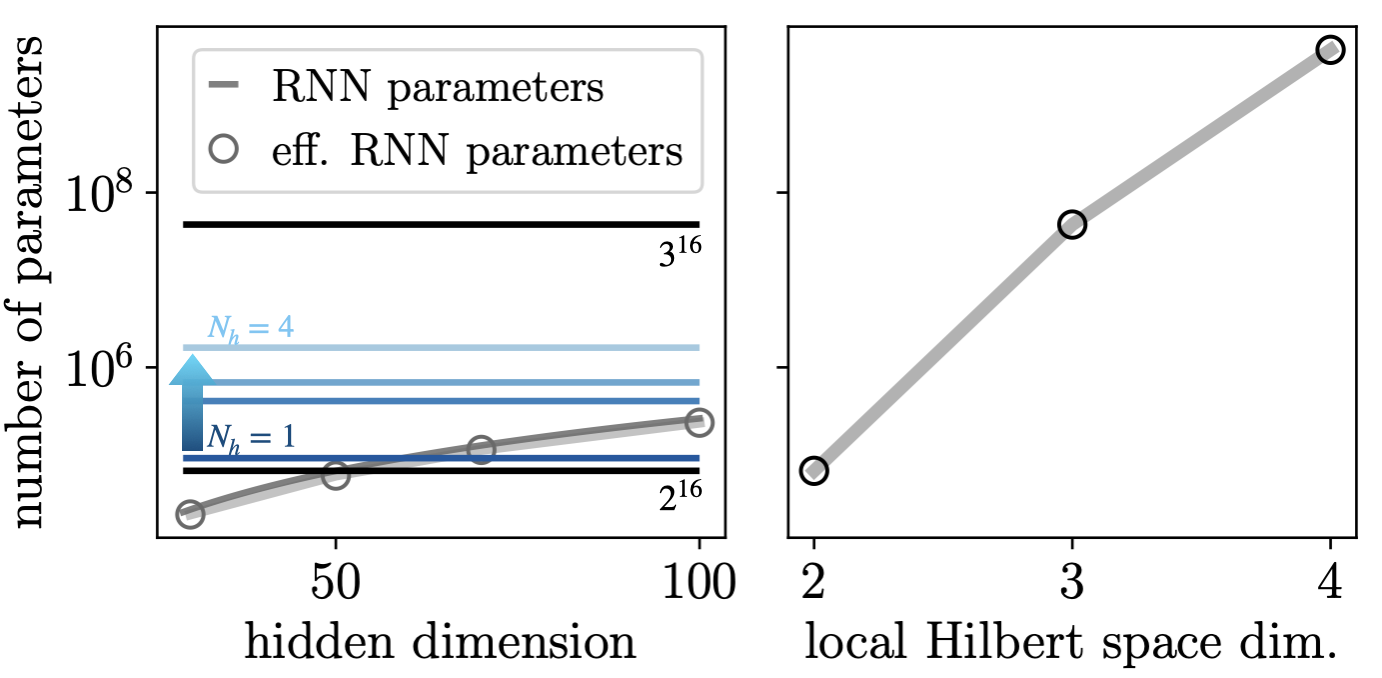}
\caption{Number of parameters for the exact wave function of a $4\times 4$ system compared to the RNN ansatz. Left: We compare the number of parameters of the exact wave function using $U(1)_{\hat{N}}\times U(1)_{\hat{S}_z}$ symmetry for $0\leq N_h \leq 4$ (blue) to the Hilbert space dimension $3^{16}$ that we want to learn with the RNN ansatz. The number of parameters of the RNN ansatz with hidden dimension $30 \leq h_d\leq 100$ is denoted in gray. Right: Hilbert space dimension for a local dimension of $2$ (Heisenberg model), $3$ ($t-J$ model) and $4$ (Fermi-Hubbard model). }
\label{fig:params}
\end{figure}

Lastly, we consider the complexity of our learning problem and compare it to the expressivity of our RNN ansatz in terms of the number of parameters that are encoded in the RNN. In Fig. \ref{fig:params} on the left, we show the number of parameters used in the RNN ansatz for the $4\times4$ $t-J$ square lattice for hidden dimensions $30\leq h_d \leq 100$. The number of parameters encoded in the ansatz is slightly lower than the number of parameters that is actually used (gray circles on the left). This is due to the way we encode the $U(1)$ symmetry in our approach, resulting in a small fraction of weights that are not updated since the respective probabilities are set to zero to obey the $U(1)$ symmetry, see Appendix \ref{appendix:symmetry}. Furthermore, we show the dimension of the Hilbert space for the same system $3^{16}$ in black, i.e. the dimension of the distribution that needs to be learned by our RNN. For the small system size that we consider in Fig. \ref{fig:params}, the Hilbert space dimension is two orders of magnitude larger than the number of RNN parameters. For the $10\times 4$ system in Fig. \ref{fig:1} however, our RNN representation has 13 orders of magnitude less parameters than the Hilbert space with dimension $3^{40}$ that is learned.

The Hilbert space dimension $3^{N_\mathrm{sites}}$ that was considered so far allows for three states per site -- spin up, down and hole --, i.e. for a variable number of holes in the system. For a fixed number of holes, the number of parameters to describe the exact state can be reduced to the Hilbert space dimension of the spin system multiplied by all combinations of how holes can be distributed on the lattice. This yields a much lower number of parameters than $3^{N_\mathrm{sites}}$, as shown by the blue lines in Fig. \ref{fig:params} for $1\leq N_h\leq 4$. In fact, for $N_h=1$ our RNNs encode even more parameters than this exact parameterization when $h_d>70$. This reveals one main problem of our RNN ansatz, namely the flexibility to encode any number of holes and hence a $3^{N_\mathrm{sites}}$-dimensional parameter space. For future studies, we envision an RNN ansatz for a fixed number of holes, reducing the dimension of the parameter space that needs to be learned and hence facilitating the learning problem.

Lastly, we would like to point out that the learning problem that we consider here is more complex than for spin systems that are typically considered with this architecture \cite{HibatAllah2020,Czischek2022,Moss2023,Roth2020}, as can be seen when comparing the Hilbert space dimensions for local dimensions $d=2$ as for spin systems, vs. $d=3$ as for the $t-J$ model in Fig. \ref{fig:params} on the right. For larger systems, this difference increases, e.g. for the $10\times 4$ system in Fig. \ref{fig:1} the Hilbert space dimension increases by seven orders of magnitude when going from a spin to a $t-J$ system (with flexible number of holes). This problem becomes even more pronounced when the Fermi-Hubbard model with local dimension $d=4$ would be considered.

\section{Summary and Outlook}
To conclude, we present a neural network architecture, based on RNNs \cite{HibatAllah2020}, to simulate ground states of the fermionic and bosonic $t-J$ model upon finite hole doping. We show that, despite many challenges due to the increased complexity of the learning problem compared to spin systems, the RNN succeeds in capturing remarkable physical properties like the shape of the dispersion, indicating the dominating emergent excitations of the systems. In order to calculate the dispersion, we present a new method that can be used with any NQS ansatz and for any lattice geometry and map out quasiparticle dispersion using the RNN ansatz for several different lattice geometries, including 1D and 2D systems. Moreover, it enables an extremely efficient calculation of dispersion relations compared to conventional methods like DMRG \cite{Vanderstraeten2015}, which usually require a time-evolution of the state \cite{Damme2021}. The dispersion scheme yields a good agreement when comparing to exact diagonalization or DMRG results, and is expected to perform even better for a better ground state convergence. In principle, it can also be combined with a translationally symmetric NQS ansatz to improve the accuracy. Furthermore, the scheme could be combined additional symmetries, e.g. rotational symmetries, enabling the calculation of $m_4$ rotational spectra \cite{Bohrdt2021}.  \\

In addition, we provide a detailed discussion on the challenges that are encountered during training our $t-J$ RNN architecture, namely $(i)$ the enlarged local Hilbert space with three states for spin up particles, spin down particles and holes, respectively, yielding $3^{N_\mathrm{sites}}$ possible configurations instead of $2^{N_\mathrm{sites}}$ as for spin systems; $(ii)$ the significant number of wave function amplitudes that are close to zero; $(iii)$ the learning plateau associated with a local minimum that is encountered for all considered optimization routines -- including annealing \cite{HibatAllah2022}, minimum-step stochastic reconfiguration (minSR) \cite{chen2023efficient} and the recently proposed SR variant based on a linear algebra trick \cite{rende2023simple} -- and the fact that SR algorithms have problems with autoregressive architectures \cite{Donatella2023}; $(iv)$ the complicated interplay between phase and amplitude optimization \cite{Bukov2021}; $(v)$ the difficulty to implement constrains on the symmetry sector under consideration, e.g. the particle number, magnetization and spatial symmetries directly into the RNN architecture \cite{Roth2020,HibatAllah2020}. Remarkably, all of these challenges are inherent to the simulation of both bosonic and fermionic systems. 
Our results indicate that the bottleneck for simulating fermionic spinful systems is the training and not the expressivity of the ansatz, and point the way to possible improvements concerning the ansatz and the training procedure.\\
\vspace{0.5cm}



\emph{Code availability.--} The code and the data used for this paper is provided here:
\url{https://github.com/HannahLange/Fermionic-RNNs/}.\\

\emph{Acknowledgements.--}
We thank Ao Chen, Ejaaz Merali, Estelle Inack, Fabian Grusdt, Lukas Vetter, Markus Heyl, Markus Schmitt, Mohammed Hibat-Allah, Moritz Reh, Roeland Wiersema, Roger Melko, Schuyler Moss, Stefan Kienle, Stefanie Czischek and Tizian Blatz for helpful and inspiring discussions. 
We acknowledge funding by the Deutsche Forschungsgemeinschaft (DFG, German Research Foundation) under Germany's Excellence Strategy -- EXC-2111 -- 390814868 and from the European Research Council (ERC) under the European Union’s Horizon 2020 research and innovation programm (Grant Agreement no 948141) — ERC Starting Grant SimUcQuam. HL acknowledges support by the International Max Planck Research School. JC acknowledges support from the Natural Sciences and Engineering Research Council (NSERC) and the Canadian Institute for Advanced Research (CIFAR) AI chair program. Resources used in preparing this research were provided, in part, by the Province of Ontario, the Government of Canada through CIFAR, and companies sponsoring the Vector Institute \url{www.vectorinstitute.ai/#partners}.

\bibliography{main.bib}

\newpage~

\newpage~

\appendix
\onecolumngrid \section*{Appendix}
\section{Recurent neural network (RNN) quantum states \label{appendix:RNNs}}
\subsection{The RNN architecture}
In the present paper we use a recurrent neural network (RNN) \cite{Hochreiter1997} to represent a quantum state defined on a 2D lattice with $N_{\mathrm{sites}}=N_x\cdot N_y$ positions occupied by $N_p\leq N_{\mathrm{sites}}$ particles, similar to Refs. \cite{HibatAllah2020, Roth2020,Czischek2022,Carrasquilla2019}.

\begin{figure}[t]
\centering
\includegraphics[width=0.48\textwidth]{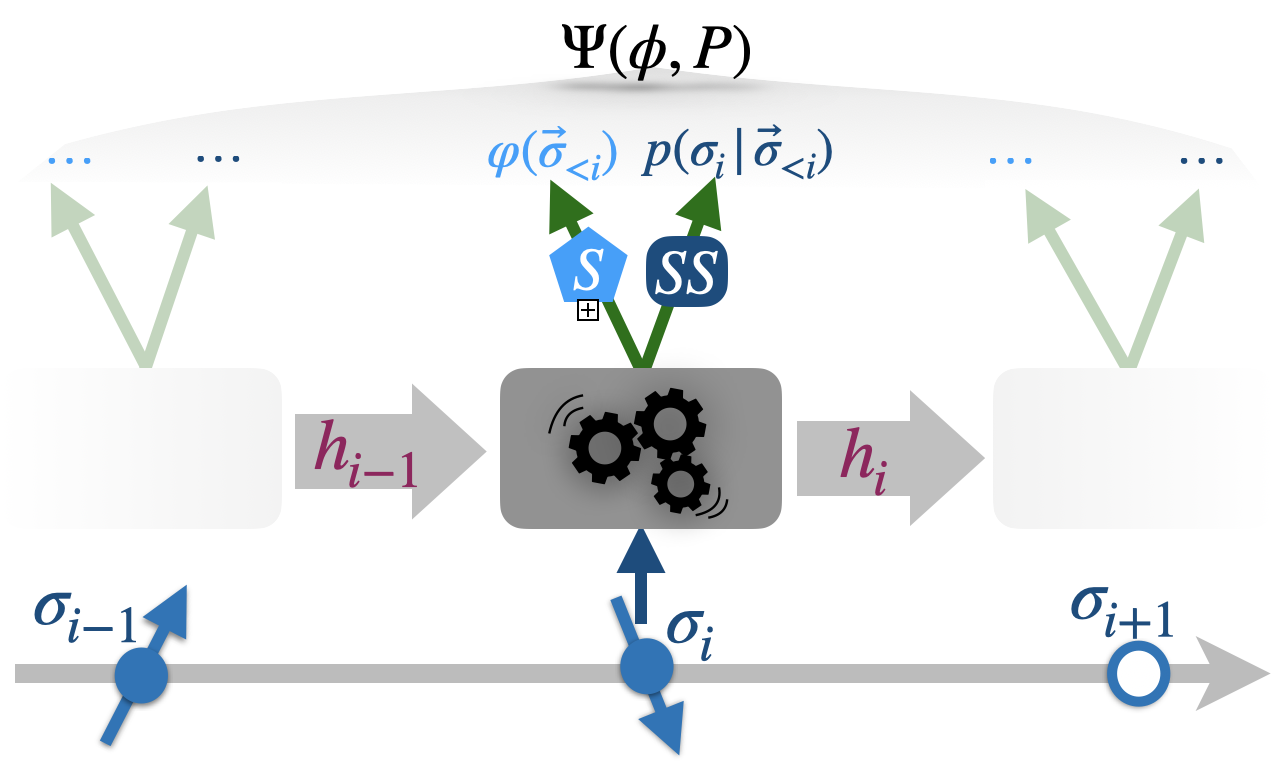}
\hspace{1.5cm }\includegraphics[width=0.3\textwidth]{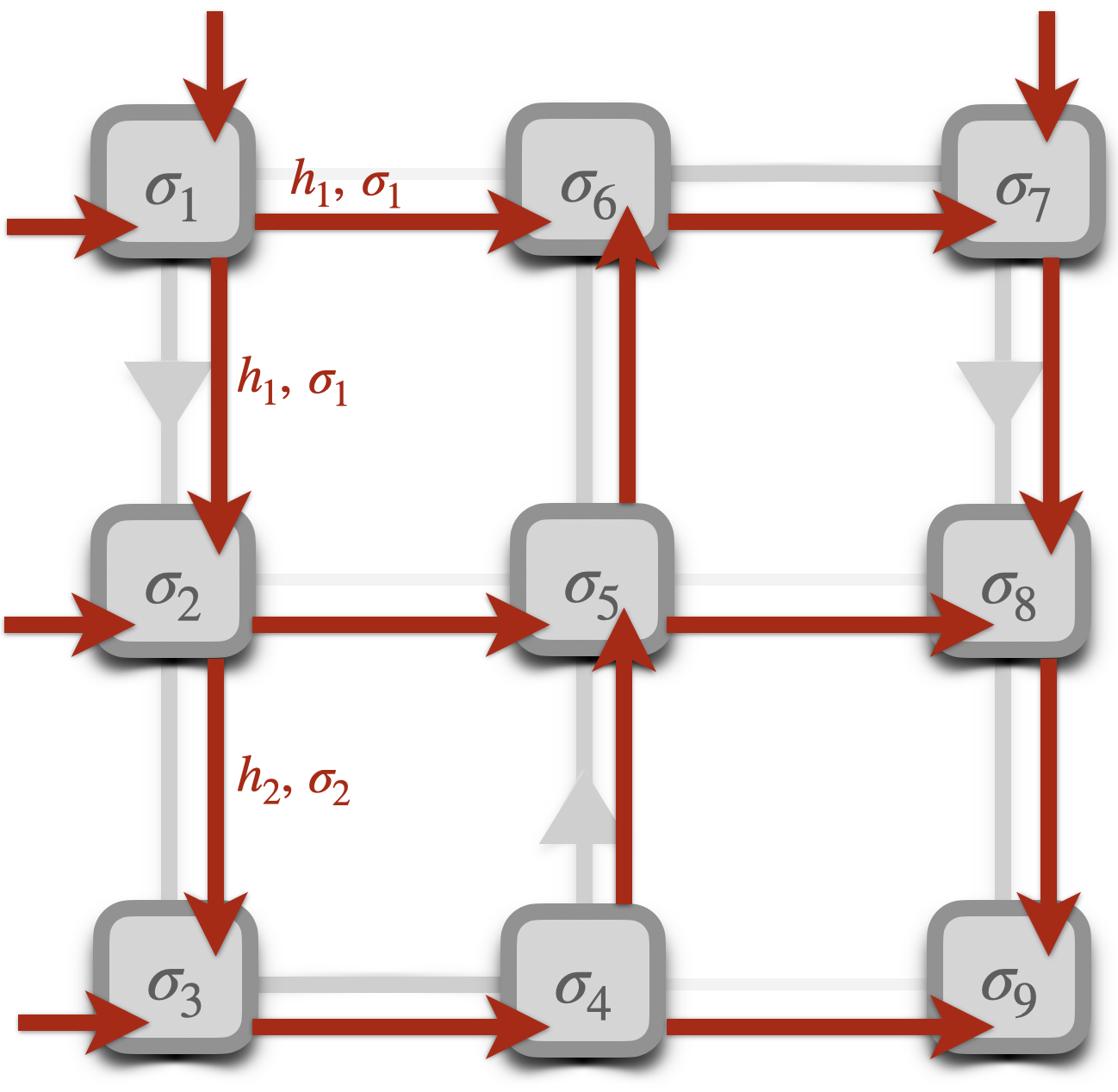}
\caption{\label{fig:RNN} Left: RNN representation of a quantum state: The hidden units $\vec{h}_i$ and the configurations $\vec{\sigma}_i$ are passed through the RNN cell for each lattice position. The network outputs the amplitude $\sqrt{P(\vec{\sigma})}$ and the phase $\phi(\vec{\sigma})$ of the RNN state representation as given by Eq. \eqref{eq:RNNstate}. Right: The two dimensional and tensorized RNN: The sampling path is represented by the dark gray lines. Internally, the hidden and configuration states, $\vec{h}_i$ and $\vec{\sigma}_i$, are passed through the network in two directions. Furthermore, inputs are passed to the tensorized version of a GRU cell (see Eq. \eqref{eq:TensorizedRNN}) at each step $i$ as proposed by Hibat-Allah et al. \cite{HibatAllah2021}.}
\end{figure}

In order to represent fermionic wave functions, we start from the same approach as for bosonic systems and use an RNN architecture consisting of $N_{\mathrm{lat}}$ (tensorized) gated recurrent units (GRUs). For one-dimensional systems, the respective 1D RNN quantum state representations is given by $N_{\mathrm{sites}}$ RNN cells and the information is passed from the first cell corresponding to the first spin of the 1D chain to the last spin in a recurrent fashion, as is shown in figure \ref{fig:RNN}. At each lattice site $i$ we define $\vec{\sigma}_i$ to denote the local spin configuration and $\vec{h}_i$ to be the so-called \enquote{hidden} state that is used to pass information from previous lattice sites through the network. Given an input $\vec{\sigma}_i\in d_v$ ($d_v$: number of features of the input data, e.g. $d_v=2$ for spin models, $d_v=3$ for the $t-J$ model and $d_v=4$ for the Fermi-Hubbard model) and a hidden state $\vec{h}_{i-1}\in d_h$, the RNN cell outputs the updated hidden state $\vec{h}_i$ as well as a conditional probability distribution and a phase, see Fig. \ref{fig:RNN}. Since it is possible to pass several sets of configurations (i.e. $N_s$ samples) through the network at once we will use the notation as vectors $\vec{\sigma}_i$ if a stack of $N_s$ configuration is considered and $\sigma_i$ for a single configuration.

 The conditional probability of finding $\sigma_i$ given a configuration $\vec{\sigma}_{<i}:=\vec{\sigma}_1\vec{\sigma}_2\dots\vec{\sigma}_{i-1}$ is given by \cite{Goodfellow2016} 
\begin{align}
P_{\vec{\lambda}}(\vec{\sigma}_i\vert \vec{\sigma}_{<i})=\vec{y}^{(1)}_i\cdot \vec{\sigma_i}
\end{align}
 with $\vec{y}^{(1)}=S(U^{(1)}\vec{h_i}+\vec{b}^{(1)})$ and the softmax activation function 
\begin{align}
S(x_i) = \frac{\mathrm{exp}(x_i)}{\sum_n \mathrm{exp}(x_n)}.
\end{align}
The total probability distribution represented by the RNN is
\begin{align}
P_{\vec{\lambda}}(\vec{\sigma}) = \Pi_i^N P_{\vec{\lambda}}(\vec{\sigma}_i\vert \vec{\sigma}_{<i}),
\end{align}
which is used to represent the amplitudes of the RNN wave function. Furthermore, since each of the conditionals is normalized, also $P_{\vec{\lambda}}(\vec{\sigma})$ is normalized, which enables very efficient sampling from the RNN wave function by going through the conditionals at each lattice site, and does not require more elaborate procedures like Monte Carlo sampling. The phase of the RNN wave function is determined by the local phases 
\begin{align}
    \phi_i(\vec{\sigma}_i\vert \vec{\sigma}_{<i}) = \pi \vec{y}^{(2)}_i\cdot\vec{\sigma}_i,
\end{align}
given by another linear layer $\vec{y}^{(2)}=\mathrm{soft}(U^{(2)}\vec{h_i}+\vec{b}^{(2)})$ and the softsign activation function 
\begin{align}
\mathrm{soft}(x_i)=\frac{x_i}{1+\vert x_i\vert}.
\end{align}
The total phase represented by the RNN is
$$
\phi_{\vec{\lambda}}(\vec{\sigma}) = \sum_i^N \phi_{\lambda,i}(\vec{\sigma}_i\vert \vec{\sigma}_{<i})
$$
and hence the full RNN wave function is given by
\begin{align}
\ket{\psi}_{\vec{\lambda}}=\sum_{\vec{\sigma}}\mathrm{exp}(i\phi_{\vec{\lambda}}(\sigma))\sqrt{P_{\vec{\lambda}}(\sigma)}\ket{\sigma}.
\end{align}

\subsection{Tensorized gated recurrent units}
In the present work we use the 2D version of RNN wave functions, as proposed in Ref. \cite{HibatAllah2021}. The underlying idea is to separate the sampling path and the information path when going through the network, as indicated in Fig. \ref{fig:RNN}. While the sampling is still done in a one-dimensional fashion (see light gray arrows), the information contained in the hidden states is passed in a 2D manner (red arrows) indicated by the blue arrows in Fig. \ref{fig:RNN}. More precisely, the hidden state of the RNN is calculated via
\begin{align}
    \vec{h}_{i,j} &= F(\left[\vec{\sigma}_{i-1,j}\vec{\sigma}_{i,j-1}\right]T_{i,j} \left[\vec{h}_{i-1,j}\vec{h}_{i,j-1}\right] +\vec{b}_{i,j})\notag \\
    &= F(W\left[\vec{\sigma}_{i-1,j}\vec{\sigma}_{i,j-1}\right]\left[\vec{h}_{i-1,j}\vec{h}_{i,j-1}\right]^T +\vec{b}_{i,j})
    \label{eq:TensorizedRNN}
\end{align}
with a nonlinear activation function $F$ and a tensor $T_{i,j}\in R^{2d_v\times2d_h \times d_h}$ or a weight matrix $W\in R^{2d_v\times2d_h \times d_h}$
Here, $\left[\vec{\sigma}_{i-1,j}\vec{\sigma}_{i,j-1}\right]$ has dimension $N_s \times 2d_v$ and $\left[\vec{h}_{i-1,j}\vec{h}_{i,j-1}\right]$ dimension $N_s \times 2 d_h$, i.e. the product of these vectors with the tensor $T_{i,j}\in R^{2d_v\times2d_h \times d_h}$ is of dimension $N_s \times d_h$ as desired.

Furthermore, we use a variant of a gated recurrent unit (GRU) instead of a simple RNN cell. GRUs tackle the difficulties of plain vanilla RNNs to capture long-term dependecies \cite{Bengio1994,Schaefer2006,Razvan2013}. In GRUs the hidden state $\vec{h}_{i,j}$ is determined by calculating
\begin{align*}
    \vec{u}_{i,j} &= \mathrm{sig}(W^1\left[\vec{\sigma}_{i-1,j}\vec{\sigma}_{i,j-1}\right]\left[\vec{h}_{i-1,j}\vec{h}_{i,j-1}\right]^T +\vec{b}^1_{i,j}) \\
    \vec{\Tilde{h}}_{i,j} &= \mathrm{tanh}(W^2\left[\vec{\sigma}_{i-1,j}\vec{\sigma}_{i,j-1}\right]\left[\vec{h}_{i-1,j}\vec{h}_{i,j-1}\right]^T +\vec{b}^2_{i,j}) \\
    \vec{h}_{i,j} &= \vec{u}_{i,j}\cdot \vec{\Tilde{h}}_{i,j}+(1-\vec{u}_{i,j})W_m\left[\vec{h}_{i-1,j}\vec{h}_{i,j-1}\right]^T, 
\end{align*}
where $W_m \in R^{2d_h\times d_h}$ is used to match the dimensions. The nonlinear activation functions \enquote{sig} and \enquote{tanh} denote the sigmoid and hyperbolic tangent activation functions respectively. In contrast to the simple RNN cell, the updated hidden state $\vec{h}_{i,j}$ is given by a combination of the previous hidden states $\vec{h}_{i-1,j}$ and $\vec{h}_{i,j-1}$ and a updated candidate $\vec{\Tilde{h}}_{i,j}$. The update gate $\vec{u}_{i,j}$ decides how much information from each of them is taken into account in the next step. This implementation is slightly different from usual implementations of GRUs which involve a so-called forget-gate and hence contain more parameters to be optimized. For the tensorized version of a full GRU cell, one would need even more parameters to match the dimensions in the forget gate which would make the optimization process very slow.

\subsection{$U(1)$ Symmetry \label{appendix:symmetry}}
Since the ground states of the $t-J$ model have conserved particle number and conserved magnetization, i.e., a $U(1)=U(1)_{\hat{N}}\times U(1)_{\hat{S}_z}$ symmetry, it is helpful to enforce this constraint on our RNN wave functions, as shown for the magnetization sector in Ref. \cite{HibatAllah2020}. The procedure that we use effectively applies a projector $\hat{P}_{\hat{S}_z=0}$ ($\hat{P}_{\hat{S}_z=0.5}$) and $\hat{P}_{\hat{N}=N_\mathrm{target}}$ for even (odd) particle numbers $N_\mathrm{target}$. This restricts the RNN wave function to the subspace of configurations under interest, yielding a simpler optimization landscape.
To satisfy our $U(1)=U(1)_{\hat{N}}\times U(1)_{\hat{S}_z}$ constrain, we utilize the following algorithm. At each site $i$, we 
\begin{enumerate}
    \item generate the RNN output $\vec{y}_i^{(1)}$ and calculate the conditional probabilities $P_{\vec{\lambda}} (\frac{1}{2}\vert\vec{\sigma}_{<i})$, $P_{\vec{\lambda}} (-\frac{1}{2}\vert\vec{\sigma}_{<i})$ and $P_{\vec{\lambda}} (0\vert\vec{\sigma}_{<i})$ for spin up, down and holes respectively.
    \item define the respective amplitudes for spin up, down and holes:
    \begin{align}
        \quad \quad a_i &= P_{\vec{\lambda}} (\vec{\sigma}_i=\frac{1}{2}\vert\vec{\sigma}_{<i}) \cdot \Theta\left(\frac{N_\mathrm{target}}{2}-N_\mathrm{up}(i) \right)\notag\\
        b_i &= P_{\vec{\lambda}} (\vec{\sigma}_i=-\frac{1}{2}\vert\vec{\sigma}_{<i}) \cdot \Theta\left(\frac{N_\mathrm{target}}{2}-N_\mathrm{down}(i) \right)\notag\\
        c_i &= P_{\vec{\lambda}} (\vec{\sigma}_i=0\vert\vec{\sigma}_{<i}) \cdot \Theta\left(N_\mathrm{target}^\mathrm{holes}-N_\mathrm{holes}(i) \right)\notag
    \end{align}
    with $N_\mathrm{target}^\mathrm{holes}=N_\mathrm{sites}-N_\mathrm{target}$, $N_\mathrm{holes}(i)=N_\mathrm{sites}-(N_\mathrm{up}(i)+N_\mathrm{down}(i))$ and $\Theta$ the heaviside function. $N_\mathrm{up}(i)$, $N_\mathrm{down}(i)$ and $N_\mathrm{holes}(i)$ are averaged values calculated from samples generated up to site $<i$.
    \item calculate the new $\Tilde{P}_{\vec{\lambda}} (\vec{\sigma}_i\vert\vec{\sigma}_{<i})$ using $a_i$, $b_i$ and $c_i$ and normalize by multipling with $\frac{1}{\sqrt{a_i^2+b_i^2+c_i^2}}$. Hence, the new probabilities $\Tilde{P}_{\vec{\lambda}}$ are also normalized.
\end{enumerate}
This procedure sets all probabilities for non-desired magnetizations and particle numbers to zero, but leaves the amplitudes of the wave function normalized to one.

\section{Optimization}
\subsection{Variational Monte Carlo (VMC)}
In order to find the ground state of the system under consideration, we use Variational Monte Carlo (VMC) \cite{Becca2017,Goodfellow2016}. VMC has already been combined in a wide range of machine learning applications (see e.g. Refs. \cite{Carrasquila2021,Melko2019}). In VMC we minimize the expectation value of the energy of the RNN trial wave function
\begin{align}
\langle E_{\vec{\lambda}}\rangle &= \frac{\langle \psi_{\vec{\lambda}} \vert \mathcal{H}\vert\psi_{\vec{\lambda}}\rangle }{\langle \psi_{\vec{\lambda}} \vert\psi_{\vec{\lambda}}\rangle}= \sum_{\vec{\sigma}}\frac{\langle \psi_{\vec{\lambda}} \vert \sigma\rangle \langle \sigma\vert\mathcal{H}\vert\psi_{\vec{\lambda}}\rangle }{\langle \psi_{\vec{\lambda}} \vert\psi_{\vec{\lambda}}\rangle}
= \sum_{\vec{\sigma}}\frac{\vert\langle \psi_{\vec{\lambda}} \vert \sigma\rangle \vert^2}{\sum_{\sigma^\prime}\vert\langle \psi_{\vec{\lambda}} \vert \sigma^\prime\rangle \vert^2 }
\frac{\langle \sigma\vert\mathcal{H}\vert\psi_{\vec{\lambda}}\rangle }{\langle \sigma \vert\psi_{\vec{\lambda}}\rangle}
= \sum_{\vec{\sigma}}P_{\vec{\lambda}}(\sigma)\, E^{\mathrm{loc}}_{\vec{\lambda}} (\sigma),
\end{align}
where we have defined the local energy
\begin{align}
    E^{\mathrm{loc}}_{\vec{\lambda}} (\sigma)=\frac{\langle \sigma\vert\mathcal{H}\vert\psi_{\vec{\lambda}}\rangle }{\langle \sigma \vert\psi_{\vec{\lambda}}\rangle}
\end{align}
and the probability distribution given by the RNN
\begin{align}
    P_{\vec{\lambda}}(\sigma) = \frac{\vert\langle \psi_{\vec{\lambda}} \vert \sigma\rangle \vert^2}{\sum_{\vec{\sigma}^\prime}\vert\langle \psi_{\vec{\lambda}} \vert \sigma^\prime\rangle \vert^2 }=\frac{\vert \psi_{\vec{\lambda}} (\sigma) \vert^2}{\sum_{\vec{\sigma}^\prime}\vert\psi_{\vec{\lambda}} (\sigma^\prime)\vert^2 }.
\end{align}

As shown e.g. in Refs. \cite{HibatAllah2020,Inui2021} one can use the cost function
\begin{align}
    \mathcal{C} = \sum_{\vec{\sigma}}\vert \psi_{\vec{\lambda}}(\sigma)\vert ^ 2 \left[  E^{\mathrm{loc}}_{\vec{\lambda}} (\sigma)-\langle E^{\mathrm{loc}}_{\vec{\lambda}}\rangle\right] 
\end{align}
to minimize both the local energy as well as the variance of the local energy. Here, $\langle E^{\mathrm{loc}}_\lambda\rangle$ is given by
\begin{align}
    \langle E^{\mathrm{loc}}_{\vec{\lambda}}\rangle = \sum_{\vec{\sigma}} \vert \psi_{\vec{\lambda}} (\sigma)\vert^2  E^{\mathrm{loc}}_{\vec{\lambda}}(\sigma)\notag
    =\frac{1}{N_s}\sum_i^{N_s}  E^{\mathrm{loc}}_{\vec{\lambda}}(\sigma_i).
\end{align}
The gradient of the cost function $C$ is given by
\begin{align}
    \partial_{{\vec{\lambda}}_i} \mathcal{C}
    &\approx\frac{2}{N_s}\mathrm{Re}\left[\sum_i^{N_s} \frac{\partial_{{\vec{\lambda}}_i} \psi^{*}_{\vec{\lambda}} (\sigma_i) }{\psi^{*}_{\vec{\lambda}} (\sigma_i)} (E^{\mathrm{loc}}_{\vec{\lambda}}(\sigma_i)-\langle E^{\mathrm{loc}}_{\vec{\lambda}}\rangle)\right] \notag \\
    &= \frac{2}{N_s}\mathrm{Re}\left[\sum_i^{N_s} \partial_{{\vec{\lambda}}_i} \mathrm{log }\psi^{*}_{\vec{\lambda}} (\sigma_i) (E^{\mathrm{loc}}_{\vec{\lambda}}(\sigma_i)-\langle E^{\mathrm{loc}}_{\vec{\lambda}}\rangle)\right].
\end{align}
The additional term in the cost function $\mathrm{Re}(\langle \partial_{{\vec{\lambda}}_i} \mathrm{log }\psi^{*}_{\vec{\lambda}} (\sigma) \rangle \langle E^{\mathrm{loc}}_{\vec{\lambda}}\rangle) = 0$ does not introduce any bias \cite{Shakir2019,HibatAllah2020}.

\subsection{Stochastic reconfiguration \label{appendix:minSR}}
\begin{figure}[b]
\centering
\includegraphics[width=0.45\textwidth]{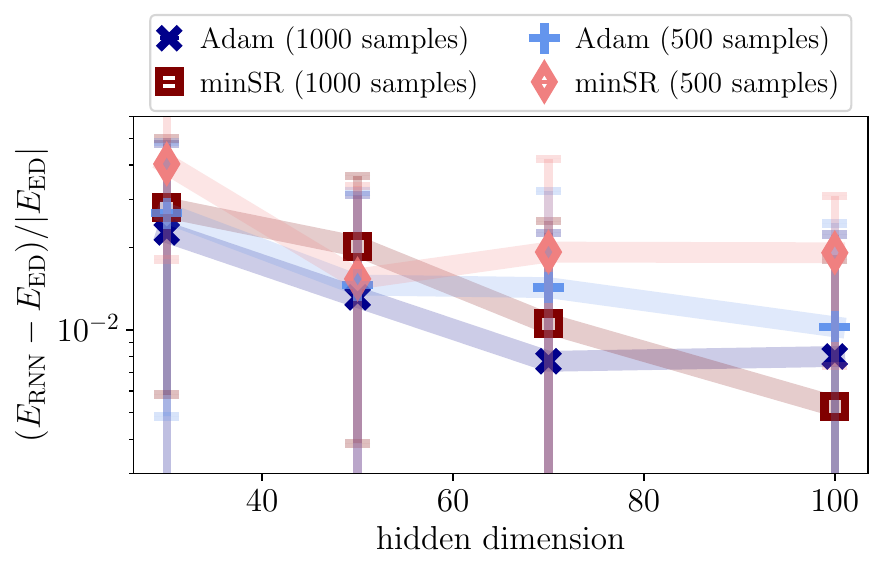}
\caption{Optimization results for the $t-J$ model for $N_h=1$ using Adam and minSR, and 500-1000 samples in each VMC step.}
\label{fig:minSR}
\end{figure}

A more elaborate approach to update the network parameters $\vec{\lambda}$ in VMC is stochastic reconfiguration (SR) \cite{Sorella1998}. It uses the local curvature of the variational manifold, measured by the quantum geometric tensor \cite{Stokes2020}.
The parameter updates are given by
\begin{align}
     \Bar{O}\delta\vec{\lambda}= \Bar{\epsilon}
\end{align}
with $\Bar{\epsilon}(\vec{\sigma}) = -\frac{1}{\sqrt{N_s}}\left( E_{\mathrm{\lambda}}(\vec{\sigma})-\langle E_{\mathrm{\lambda}}(\vec{\sigma}) \rangle\right)$ and $O_{\vec{\sigma} k} := \partial\vec{\lambda}_k \mathrm{\psi}_{\vec{\lambda}} (\vec{\sigma}) $, $\Bar{O}_{\vec{\sigma} k} = \frac{1}{\sqrt{N_s}} \left( O_{\vec{\sigma} k} -\langle O_{\vec{\sigma} k} \rangle \right)$ and $S=\Bar{O}^\dagger\Bar{O}\in \mathbb{C}^{N_{\vec{\lambda}}\times N_{\vec{\lambda}}}$, where $N_{\vec{\lambda}}$ is the number of network parameters.
In conventional SR, this equation is solved by multiplying $S^{-1}\Bar{O}^\dagger$ from the left, yielding the update rule
\begin{align}
    \delta\vec{\lambda} = S^{-1} \Bar{O}^\dagger \Bar{\epsilon}\, .
\end{align}
Hereby, the $S$ matrix -- a matrix of dimension $N_{\vec{\lambda}}\times N_{\vec{\lambda}}$ -- has to be inverted, which becomes computationally costly for large numbers of parameters $N_{\vec{\lambda}}$ as in our case. Therefore, we use the recently presented minimum-step SR (minsR) algorithm \cite{chen2023efficient} or the SR variant based on a linear algebra trick by Rende et al. \cite{rende2023simple}. 

\subsubsection{Minimum-step step SR (minSR)}
In minSR, Eq. \eqref{eq:SR} is solved by defining $T:=\Bar{O}\Bar{O}^\dagger \in \mathbb{C}^{N_s\times N_s}$ and using the identity $1=\Bar{O}^\dagger (\Bar{O}^\dagger)^{-1}$, resulting in
\begin{align}
    \Bar{\epsilon} &= \Bar{O}\delta\vec{\lambda}= \Bar{O}\Bar{O}^\dagger (\Bar{O}^\dagger)^{-1}\delta\vec{\lambda}=T (\Bar{O}^\dagger)^{-1}\delta\vec{\lambda}\notag \\
    &\Leftrightarrow \delta\vec{\lambda} = \Bar{O}^\dagger T^{-1} \Bar{\epsilon}\, .
\end{align}
In Ref. \cite{chen2023efficient} in was shown that this variant of SR can achieve extremely high accuracies with CNNs. However, as shown in Fig. \ref{fig:minSR}, in our case the minSR are not systematically better than the results obtained with Adam.

\subsubsection{SR variant by Rende et al.}
A recently proposed method by Rende et al. \cite{rende2023simple} rewrites $S$ as
\begin{align}
    S = \mathrm{Re}\, \Bar{O} \mathrm{Re}\, \Bar{O}^T + \mathrm{Im}\, \Bar{O} \mathrm{Im}\, \Bar{O}^T = XX^T
\end{align}
with $X=\mathrm{Concat}(\mathrm{Re}\,\Bar{O}, \mathrm{Im}\,\Bar{O} )\in \mathbb{R}^{N_{\vec{\lambda}}\times 2N_s}$. Furthermore, when defining $\vec{f}_{\vec{\sigma}}=\mathrm{Concat}(\mathrm{Re}\,\Bar{\epsilon}(\vec{\sigma}), -\mathrm{Im}\,\Bar{\epsilon}(\vec{\sigma}) ) \in \mathbb{R}^{2N_s}$, the SR update rule becomes
\begin{align}
    \delta {\vec{\lambda}}_k = (XX^T)_{k k^\prime} X_{k^\prime \vec{\sigma}}\vec{f}_{\vec{\sigma}}\,,
\end{align}
and with a linear algebra identity Eq. \eqref{eq:rende} follows.

\subsection{Number of parameters}

\begin{figure}[htp]
\centering
\includegraphics[width=0.48\textwidth]{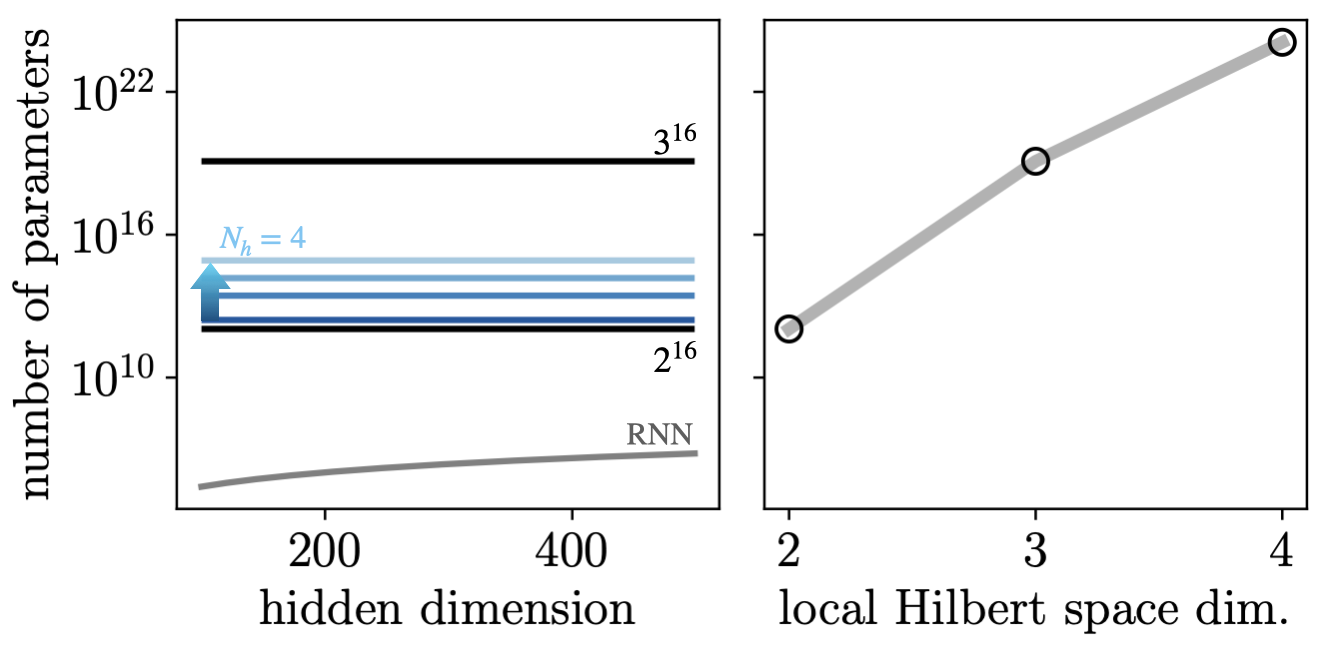}
\caption{Number of parameters for the exact wave function of a $10\times 4$ system compared to the RNN ansatz. Left: We compare the number of parameters of the exact wave function using $U(1)_{\hat{N}}\times U(1)_{\hat{S}_z}$ symmetry for $0\leq N_h \leq 4$ (blue) to the Hilbert space dimension $3^{40}$ that we want to learn with the RNN ansatz. The number of parameters of the RNN ansatz with hidden dimension $h_d$ is denoted in gray. Right: Hilbert space dimension for a local dimension of $2$ (Heisenberg model), $3$ ($t-J$ model) and $4$ (Fermi-Hubbard model). }
\label{fig:params2}
\end{figure}

In the main text, see Sec. \ref{sec:params}, we discuss the number of parameters needed for a exact parameterization of a $4\times4$ $t-J$ system, compared to the number of RNN parameters used in the RNN ansatz. The same comparison for ground state on the $10\times 4$ square lattice, see Fig. \ref{fig:1}, are shown in Fig. \ref{fig:params2}. It can be seen that in this case the number of RNN parameters is several orders of magnitude smaller than all exact parameterizations of the $t-J$ model ground states. However, also the difference between a parameterization using a fixed number of holes (shown for $1\leq N_h \leq 4$) and of a state with any number of holes, $3^{40}$ is much larger than for the smaller system. This reveals one of the problems of our architecture, namely the flexibility to encode any number of particles and any magnetization in principle. Instead, the amplitudes with undesired particle number and magnetization are only set to zero during the sampling process, as explained in Appendix \ref{appendix:RNNs}.

\section{RNN dispersion relations \label{appendix:RNNdispersion}}

For all calculated dispersion relations in the main text, we show averages over the last $100$ training iterations, each with $200$ samples, with the respective error bars as shown in Figs. \ref{fig:1} to \ref{fig:2x9_triangular}. The training is stopped when the momentum is close to the target momentum over $500$ to $5000$ training iterations, depending on the state under consideration.

\subsection{Translational invariance}
\begin{figure}[t]
\centering
\includegraphics[width=0.45\textwidth]{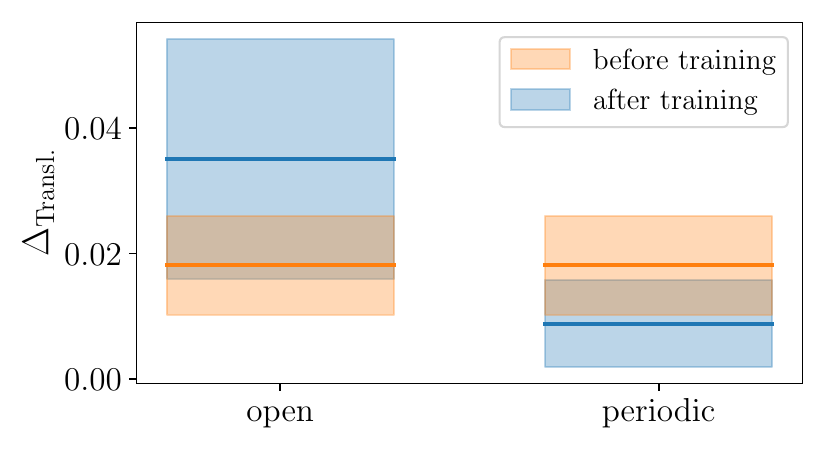}
\hspace{0.2cm}
\includegraphics[width=0.5\textwidth]{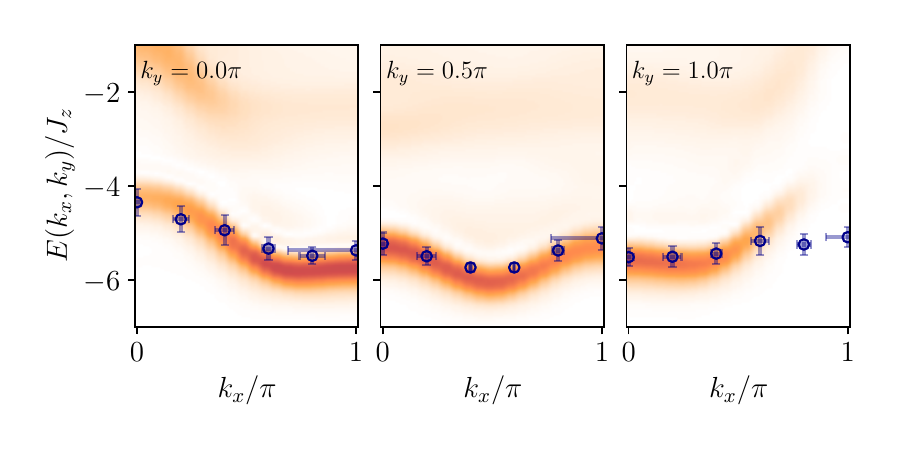}
\caption{Left: The difference between RNN log-probabilities for samples $\vec{\sigma}$ and $\hat{T}_{\vec{R}} \vec{\sigma}$, $\Delta_\mathrm{Transl.}$, here with $\vec{R}=2a\vec{e}_x$ and $\vec{R}=2a\vec{e}_y$, for a $4\times 4$ $t-J$ system with open and periodic boundaries and before (orange) and after (blue) the training. Right: Quasiparticle dispersion for a single hole on a $t-J$ square lattice with $10\times 4$ sites and open boundaries in $x$, periodic boundaries in $y$ direction, obtained from the RNN (blue) and  compared to the MPS spectral function of Ref. \cite{Bohrdt2020}, with the MPS spectral weight given by the colormap. At $k_y=\pi$ and $k_x>0.8\pi$, the spectral weight is suppressed, causing problems for our RNN scheme.}
\label{fig:dispersions}
\end{figure}

As explained in the main text, the method to calculate dispersions from NQS relies on the fact that samples drawn from the NQS are approximately translational invariant. Fig. \ref{fig:dispersions} compares the difference of NQS log-probabilities for samples $\vec{\sigma}$ and $\hat{T}_{\vec{R}} \vec{\sigma}$,
\begin{align}
    \Delta_\mathrm{Transl.} = \sum_\mu \frac{1}{N_s}\sum_i \frac{(\mathrm{log}\vert \psi_{\vec{\lambda}}(\sigma_i)\vert^2-\mathrm{log}\vert \psi_{\vec{\lambda}}(\hat{T}_{\bf{e}_\mu}\sigma_i)\vert^2)^2}{(\mathrm{log}\vert \psi_{\vec{\lambda}}(\sigma_i)\vert^2+\mathrm{log}\vert \psi_{\vec{\lambda}}(\hat{T}_{\bf{e}_\mu}\sigma_i)\vert^2)^2}\, ,
\end{align}
for an open and periodic $4\times 4$ system before and after the training. As expected, $\Delta_\mathrm{Transl.}$ is lower for the periodic, translational invariant case than for the open system. Furthermore, the translational invariance decreases compared to the initial random initailization for the periodic system.

\subsection{The effect of suppressed spectral weight}
Another remark in the main text concerns the MPS spectrum of the $10\times 4$ system in Fig. \ref{fig:1}. There is a small region of suppressed spectral weight near at $(k_x>0.8\pi,k_y=\pi)$ in the MPS spectral function of the $t-J$ system \cite{Bohrdt2018}, a region of momenta that is not shown in Fig. \ref{fig:1} but in Fig. \ref{fig:dispersions}. In Ref. \cite{Bohrdt2018} it is discussed that this feature has a strong $t/J$ dependence, in agreement with the parton picture of the polaron \cite{Grusdt2018}, with vanishing suppression for $t\leq J$, but that actually states are expected in this regime near $(\pi,\pi)$. 

The vanishing spectral weight indicates the fact that the state near $(\pi,\pi)$ has a vanishingly small overlap with the ground state at $(\pi/2,\pi/2)$. This causes problems for the NQS dispersion scheme since the momentum training is started from the ground state. As shown in Fig. \ref{fig:dispersions}, the suppression indeed coincides with a regime where the NQS scheme has problems with learning the correct low-energy state.

\end{document}